\documentclass[aps,prx,twocolumn,superscriptaddress,letterpaper]{revtex4-1}

\usepackage{amssymb}
\usepackage{amsmath}
\usepackage{bm}
\usepackage{graphicx}
\usepackage{color}
\usepackage{epsfig}

\begin{document}

\renewcommand{\figurename}{Fig.}

\title{Observation of Protected Photonic Edge States Induced By\\Real-Space Topological Lattice Defects}

%\date{\today}

\author{Qiang Wang}
\affiliation{Division of Physics and Applied Physics, School of Physical and Mathematical Sciences, Nanyang Technological University,
Singapore 637371, Singapore}

\author{Haoran Xue}
\affiliation{Division of Physics and Applied Physics, School of Physical and Mathematical Sciences, Nanyang Technological University,
Singapore 637371, Singapore}

\author{Baile Zhang}
\email{blzhang@ntu.edu.sg}

\affiliation{Division of Physics and Applied Physics, School of Physical and Mathematical Sciences, Nanyang Technological University,
Singapore 637371, Singapore}
\affiliation{Centre for Disruptive Photonic Technologies, Nanyang Technological University, Singapore, 637371, Singapore}

\author{Y. D. Chong}
\email{yidong@ntu.edu.sg}

\affiliation{Division of Physics and Applied Physics, School of Physical and Mathematical Sciences, Nanyang Technological University,
Singapore 637371, Singapore}

\affiliation{Centre for Disruptive Photonic Technologies, Nanyang Technological University, Singapore, 637371, Singapore}

\begin{abstract}
Topological defects (TDs) in crystal lattices are elementary lattice imperfections that cannot be removed by local perturbations, due to their real space topology.  We show that adding TDs into a valley photonic crystal generates a lattice disclination that acts like a domain wall and hosts topological edge states.  The disclination functions as a freeform waveguide connecting a pair of TDs of opposite topological charge.  This interplay between the real-space topology of lattice defects and band topology provides a novel scheme to implement large-scale photonic structures with complex arrangements of robust topological waveguides and resonators.
\end{abstract}

%\pacs{...}

\maketitle

\section{Introduction}

The field of topological photonics \cite{Ozawa2019}, which has emerged over the past decade, seeks to use ideas from topological band theory \cite{Bansil2016} to realize photonic modes that are protected against various forms of disorder.  Possible applications for this new class of photonic devices are still being explored, and may include robust waveguides and delay lines \cite{Hafezi2011, Wang2008, Wang2009, Dong2017, Shalaev2019}, frequency converters \cite {Hadad2018, wang2019}, and lasers \cite{StJean2017, Zhao_Feng2018, Parto2018, Ota2018, Bandres2018, Harari2018}.  A key challenge to finding practical uses for topological photonic modes is that they are typically only robust against specific types of disorder.  For example, photonic crystals based on valley Hall insulators \cite{Ma2016, Dong2017, Gao2017, He2019} and topological crystalline insulators \cite{WuHu2015, Barik2018} have been intensively studied, particularly in the nanophotonic regime, because they can be implemented using ordinary dielectric or metallic materials.  Since their topological features are tied to the presence of an underlying lattice symmetry, their topological edge states are only protected against backscattering in certain configurations (e.g., 120$^\circ$ bends in valley photonic crystal edges \cite{Ma2016}), whereas other configurations can induce backscattering and mode localization.

Topological defects (TDs) in crystal lattices are elementary lattice imperfections that cannot be removed by local perturbations, due to their real space topology \cite{Mermin1979}.  In condensed matter systems, TDs are responsible for many interesting effects, including acting as seeds of disorder in the melting of two-dimensional (2D) solids \cite{Kosterlitz2017}.  Honeycomb lattices, such as graphene, host a particularly notable class of TDs consisting of five- and seven-membered rings, which act upon 2D Dirac cone states like singular matrix-valued gauge fields carrying $\pi/2$ magnetic flux \cite{Gonzalez1993,  Lammert2000, Vozmediano2010, Kotakoski2011,deSouza2014}.  Experimental evidence of these electronic features has proven difficult to obtain, due in part to challenges in sample preparation \cite{Yazyev2010, Kotakoski2011, Lahiri2010, Huang2011, Warner2012}.  Such difficulties can be overcome by photonic structures, which can realize many phenomena that are hard to observe in condensed matter settings.  For example, photonic honeycomb lattices (``photonic graphene'') have been shown to exhibit unconventional edge states that are difficult to stabilize in real graphene \cite{Plotnik2014}.  The aforementioned physical effects of TDs, however, have yet to be explored on photonic platforms.

Here, we present a theoretical and experimental study of topologically protected waveguiding aided by TDs in valley photonic crystals (VPCs).  In honeycomb lattices with broken sublattice symmetry, TDs are the termination points of disclinations---string-like lattice defects---that cannot be gauged away \cite{Ruegg2013, deSouza2014}. We show that these disclinations are locally equivalent to domain walls of Valley Hall insulators \cite{Ma2016, Dong2017, Gao2017} and thus function as robust topological edge state waveguides \cite{Ozawa2019}.

TDs supply a novel and interesting relationship between topological features and edge states that is completely different from the usual bulk-edge correspondence principle.  They carry topological charges in real space (negative for pentagonal TDs, positive for heptagonal TDs) that stem from the configuration of the lattice instead of bandstructure features defined in momentum space.  Positive and negative TDs are joined pairwise by disclinations hosting edge states; but unlike standard VPC domain walls, one can go smoothly from one domain to the other by encircling a TD without crossing the disclination.  This feature requires the presence of TDs and does not occur in perfectly crystalline VPCs.  The disclinations can follow curved paths and are not restricted to any global axes \cite{Ma2016, Gao2017}.  They are also not limited to forming loops or ending at external boundaries, and can form open arcs (bounded by the TDs) that act as one-dimensional (1D) Fabry-P\'erot resonators based on counterpropagating topological edge states.

Previous research has shown that amorphous photonic structures (which possess short-range positional order without long-range order) can exhibit isotropic photonic band gaps, and can be inscribed with freeform curved waveguides \cite{Miyazaki2003, Edagawa2008, Florescu2009, Yang2010, Imagawa2010, Man2013, Florescu2013}.  A key limitation of such waveguides is that, as 1D disordered transport channels, they are highly susceptible to Anderson localization \cite{Kramer1993}, making it necessary to perform structural fine-tuning to optimize the localization lengths of the waveguide modes \cite{Florescu2013}.  We show that the present disclination-based waveguides display much greater resistance against Anderson localization, without structural fine-tuning, thanks to their connection to Valley Hall edge states \cite{Ozawa2019, Ma2016}.  Smooth curves in the waveguide act like large-scale impurities, while sharper bends can be implemented using local 120$^\circ$ bends, both of which induce negligible inter-valley scattering and hence preserve the topological protection of the edge states \cite{Ma2016}.  Unlike earlier demonstrations of topological waveguiding in amorphous or quasicrystalline lattices \cite{Bandres2016, Mitchell2018}, the present VPC-based design does not require breaking time-reversal symmetry.

\section{Topological Defects in Photonic Lattices}
\label{sec:structure}

\begin{figure}
\centering
\includegraphics[width=\columnwidth]{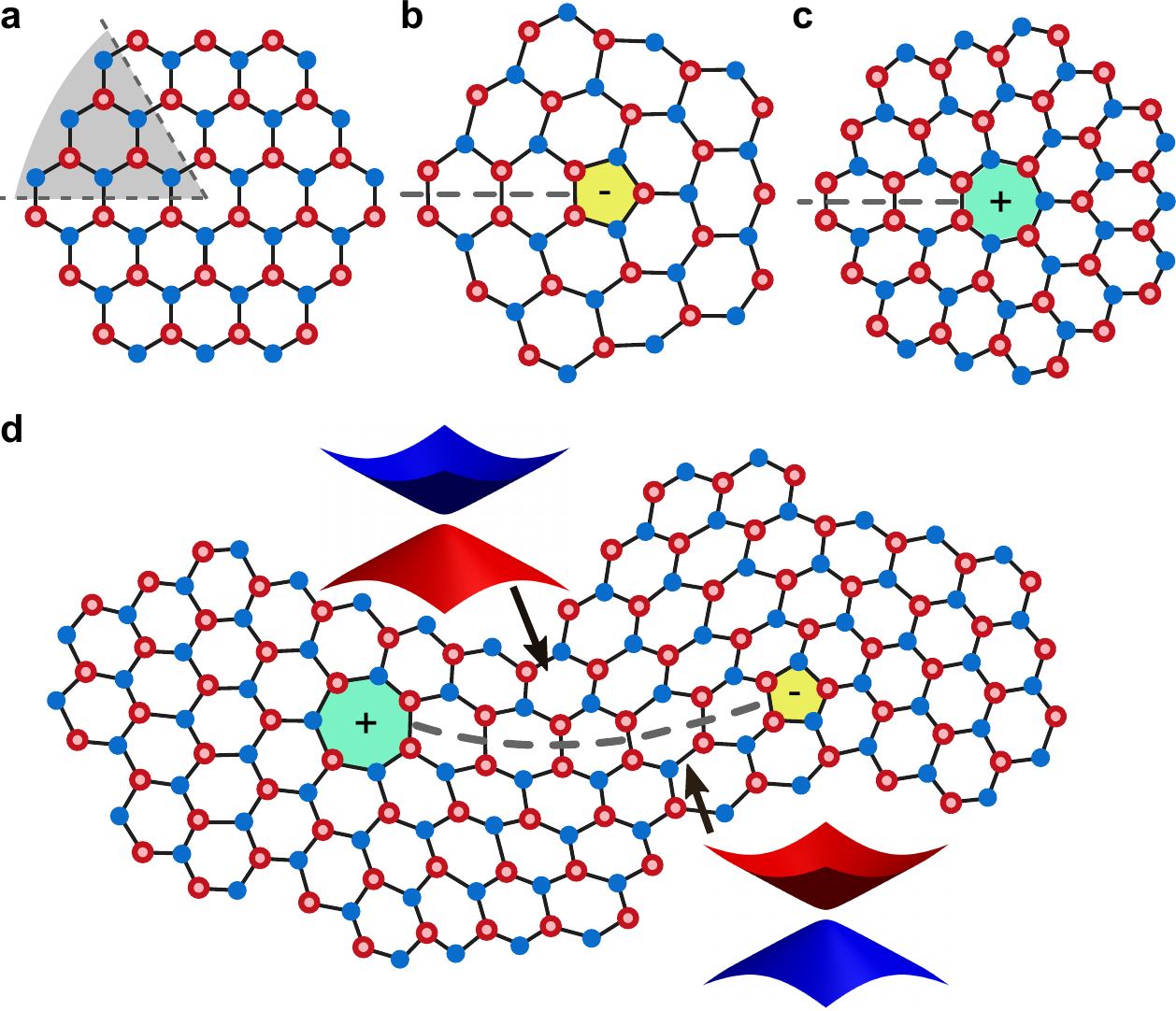}
\caption{Schematic of valley photonic crystal  lattices with TDs.  (a), Defect-free honeycomb lattice with red (blue) circles denoting sites on sublattice $A$ ($B$), and black lines indicating nearest neighbor bonds.  The gray region bounded by dashes indicates a $\pi/3$ sector whose deletion generates a pentagonal TD. (b--c), Lattice containing a pentagonal TD (labeled -) or heptagonal TD (labeled +) and a disclination (dashes) comprising a string of bonds joining sites on the same sublattice.  (d), Lattice with one pentagonal TD and one heptagonal TD, connected by a disclination forming an open arc.}
\label{fig:structure}
\end{figure}

Consider the honeycomb lattice shown in Fig.~\ref{fig:structure}(a).  The red and blue circles represent the $A$ and $B$ sublattices, which have on-site masses $m_{A} = -m_{B} = m$. As shown in Fig.~\ref{fig:structure}(b), by deleting the $\pi/3$ sector marked by the dashed lines and reattaching the seams, we can generate a pentagonal TD, which features a five-membered ring \cite{Gonzalez1993}.  Likewise, inserting an additional $\pi/3$ sector yields a heptagonal TD, as shown in Fig.~\ref{fig:structure}(c).  Each topological defect is attached to a string disclination, marked by the gray dashes in Fig.~\ref{fig:structure}(b)--(c).  On opposite sides of the disclination, sites of the same sublattice are nearest neighbors, whereas all other nearest neighbor pairs occupy different sublattices.

When $m = 0$, the disclination is ficticious in the sense that it can be moved around freely, so long as it terminates at the topological defect, by adjusting the assignment of sites to the $A$ and $B$ sublattices.  In the continuum limit, the states of the honeycomb lattice are described by a pair of Dirac cones.  The introduction of a TD imposes a disclination associated with a nontrivial boundary condition for the Dirac cone states, which several authors have shown can be gauged away---i.e., the boundary condition can be transformed into a regular continuity condition via the introduction of a matrix-valued gauge field \cite{Gonzalez1993,Lammert2000,Lammert2004,Ruegg2013}.

For $m \ne 0$, the case we are primarily interested in, the disclination is physical and cannot be gauged away \cite{Ruegg2013}.  The disclination consists of neighboring sites of the same sublattice, which is strongly reminiscent of valley Hall domain walls and hence indicates that valley Hall-like topological edge states should exist along the disclination.  (But unlike a valley Hall domain wall, which is only allowed to form a loop or terminate at exterior lattice edges, the disclination can terminate at a point in the bulk, at the position of the topological defect.)  In Appendix~\ref{sec:dirac-cone-appendix}, we present a theoretical analysis of the Dirac cone states in the continuum limit, showing that they indeed experience a valley Hall-like domain wall at the disclination---i.e., each valley sees a change in sign of the Dirac mass across the disclination.

\begin{figure*}
\centering
\includegraphics[width=\textwidth]{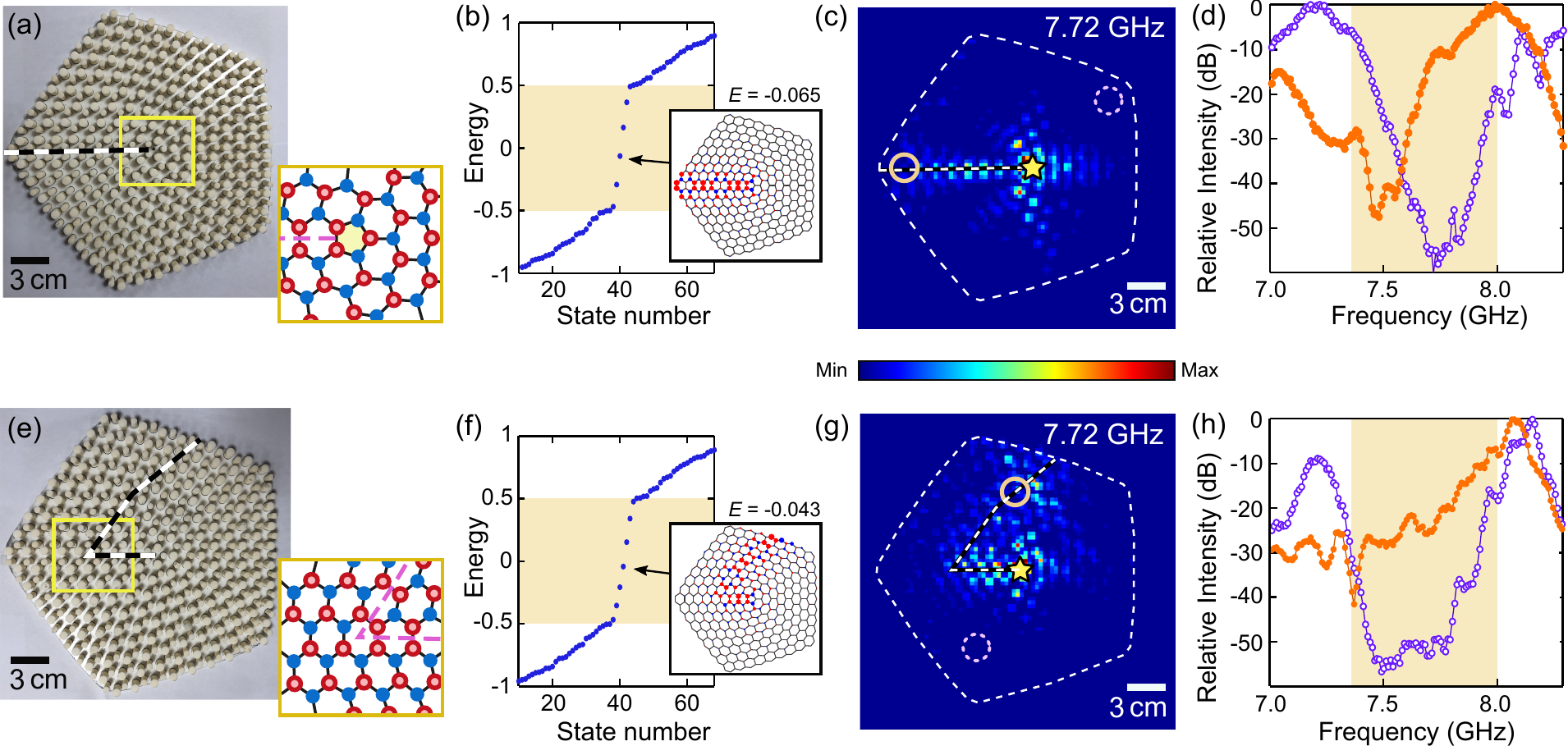}
\caption{Experimental and numerical studies of photonic lattices with a pentagonal TD.  (a) Photograph of the experimental sample with top plate removed.  Dielectric pillars are arranged in a lattice containing a TD (the associated disclination is marked by black and white dashes), and the region outside the lattice is an empty parallel plate waveguide.  The structure parameters are described in Appendix~\ref{sec:design}.  Inset: schematic of a portion of the lattice indicated by a yellow box, with red and blue circles representing $A$ and $B$ sites.  (b) Energy spectrum of a tight-binding model with the given lattice configuration, nearest neighbor hopping $t = 1$, and on-site mass $m_{A/B}= \pm {0.5}$.  The bulk band gap of an equivalent crystalline honeycomb lattice (a valley Hall insulator) is indicated in yellow.  Inset: spatial distribution of a typical eigenstate in the band gap, with red (blue) circles representing wavefunction magnitudes on $A$ ($B$) sites.  (c) Experimentally measured distribution of the electric field intensity, $|E_z|^2$, for a dipole source located at the topological defect (yellow star).  White dashes indicate the sample boundary.  The scanning resolution is $3\,\textrm{mm} \times 3\,\textrm{mm}$.  (d) Frequency dependence of $|E_z|^2$ averaged over two regions of the sample, close to the disclination (orange points) and away from the disclination (purple points).  These regions are indicated by the solid and dashed circles in (c), respectively.  The bulk band gap is indicated in yellow.  Each data set is normalized to its maximum value in the given frequency range.  (e)--(h) The corresponding results for a sample with a disclination containing a sharp bend.}
\label{fig:experiment_pentagon}
\end{figure*}

To confirm the existence of edge states bound to disclinations, we performed experiments on VPC with dielectric pillars.  The pillars are sandwiched between parallel metal plates, and the experiments are performed in the microwave regime.  Details of the setup are given in Appendix~\ref{sec:design}.

In the first set of experiments, we designed and fabricated VPC with a single pentagonal TD at the center, as shown in Fig.~\ref{fig:experiment_pentagon}(a) and (e).  To create these patterns, we start with a $2\pi/5$ or $2\pi/7$ sector, populate it with an unoptimized triangular lattice, and close-pack those lattice sites using the molecular dynamics simulator LAMMPS \cite{lammps}.  The other sectors are then determined by five- or seven-fold rotational symmetry, yielding a lattice centered on a TD.  The honeycomb-like dual lattice is generated by interpreting the triangular lattice sites as the incenters of the honeycomb motifs \cite{Mansha2016}.  

The lattice sites are ``colored'' by assigning them to sublattice $A$ or $B$, corresponding to pillar radius $r_a$ or $r_b$ respectively (see Appendix~\ref{sec:design}).  Wherever possible, neighboring sites are assigned to opposite sublattices, but this cannot be achieved everywhere: there is always a disclination---a string of lattice edges joining neighboring sites of the same sublattice---emanating from the topological defect.  By changing the lattice coloring, we can vary the path of the disclination, including making it turn sharp corners.  The top and bottom rows of Fig.~\ref{fig:experiment_pentagon} show results for two different disclination choices.

To study the qualitative features of these lattice configurations, we calculate the energy spectrum of the corresponding tight-binding models. Fig.~\ref{fig:experiment_pentagon}(b) and (f) show results for the two different disclinations, with nearest neighbor hopping $t = 1$ and on-site mass $m_A = 1/2$ and $m_B = -1/2$ for the $A$ and $B$ sublattices.  For these parameters (which are chosen for convenience, not to fit experiments), the equivalent honeycomb lattice has a bulk band gap at $m_B < E < m_A$.  The numerical results show that the lattices with a TD have eigenstates in the band gap.  As shown in the insets of Fig.~\ref{fig:experiment_pentagon}(b) and (f), the states are localized to the disclination, similar to domain wall states of valley Hall insulators.

Fig.~\ref{fig:experiment_pentagon}(c) and (g) plot the experimentally measured intensity profiles for transverse mangetic (TM) waves emitted by a dipole source placed at the TD (oriented perpendicular to the plane).  The excitation frequency of $7.72\,\textrm{GHz}$ lies within the bulk band gap of an equivalent VPCl with the same choice of parameters ($7.36\,\mathrm{GHz}$--$8.04\,\textrm{GHz}$; see Appendix~\ref{sec:design}).  We observe that the light emitted by the dipole is guided along the disclination, including around a sharp corner in the case of Fig.~\ref{fig:experiment_pentagon}(g).  In Fig.~\ref{fig:experiment_pentagon}(d) and (h), we plot the experimentally measured frequency dependence of the local field intensities averaged over different regions of the lattice.  In the frequency range of the bulk band gap, there is a clear intensity dip when the sampling region lies away from the disclination, but no dip when the sampling region lies over the disclination.

For comparison, we also investigated a ``photonic graphene'' lattice in which every pillar has the same radius (i.e., sublattice symmetry is unbroken).  According to previous theoretical studies,TDs in such lattices should not create localized resonances \cite{Lammert2000, Lammert2004, Ruegg2013}, and due to the lack of sublattice symmetry breaking there is no disclination on which edge states can appear.  This is consistent with our experimental findings, which are summarized in Appendix~\ref{sec:graphene}.

\section{ Waveguiding with topological and non-topological defects}

\begin{figure}
\centering
\includegraphics[width=\columnwidth]{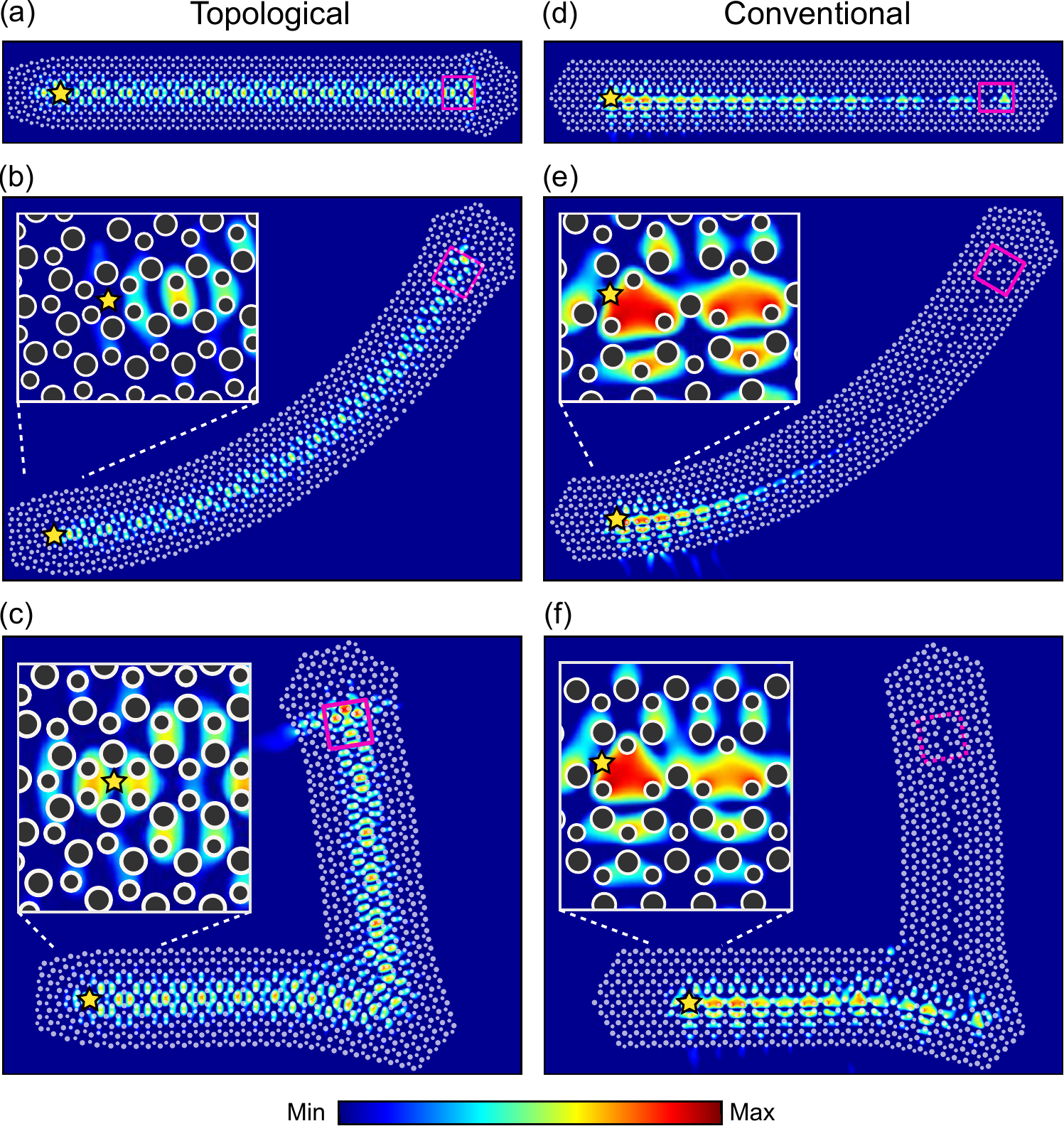}
\caption{Comparison of topological and non-topological edge states in VPCs. (a)--(c) Simulated field distributions for topological waveguides.  The pillars are indicated by white circles.  Each waveguide runs between two TDs along a path that is (a) straight, (b) curved, or (c) sharply bent.  A $6.98\,\textrm{GHz}$ dipole source is placed near one of the TDs (yellow star).  The heat map colors correspond to $\log |E_z|^2$ (a logarithmic scale is used so that the intensities in the non-topological case are more visible), with all three plots normalized to the maximum intensity along the straight waveguide (a).  (d)--(f) Simulated field distributions of $\log |E_z|^2$ for non-topological waveguides generated by removing pillars from a lattice that does not contain TDs.  A $6.98\,\textrm{GHz}$ dipole source is placed near one end of the waveguide (yellow star).  All three plots are normalized to the maximum intensity along the straight waveguide (d).  In all subplots, the  structure parameters are as stated in Appendix~\ref{sec:design}.}
\label{fig:comparison}
\end{figure}

The lattices studied in the previous section have a relatively simple arrangement, with five- or seven-fold rotational symmetry around a central TD.  In this section, we study more complex VPC with TDs hosting freeform topological waveguides.

%There have been many earlier studies on waveguides embedded in amorphous photonic structures \cite{Miyazaki2003, Florescu2009, Florescu2013}.  Light is confined in such waveguides by the existence of a complete band gap in the bulk, while the non-crystalline nature of the lattice allows the waveguides to follow curved routes with arbitrary bend angles (unlike photonic crystal waveguides, which are restricted to bend angles matching the high-symmetry directions of the crystal).  One key limitation of such waveguides is that, as 1D disordered transport channels, they are highly susceptible to Anderson localization \cite{Kramer1993}.  While it is possible to optimize their localization lengths by fine-tuning structural parameters along the waveguide \cite{Florescu2013}, that nontrivial step seems to diminish the appealing simplicity of using amorphous structures to host freeform waveguides.

\begin{figure}
  \centering
  \includegraphics[width=\columnwidth]{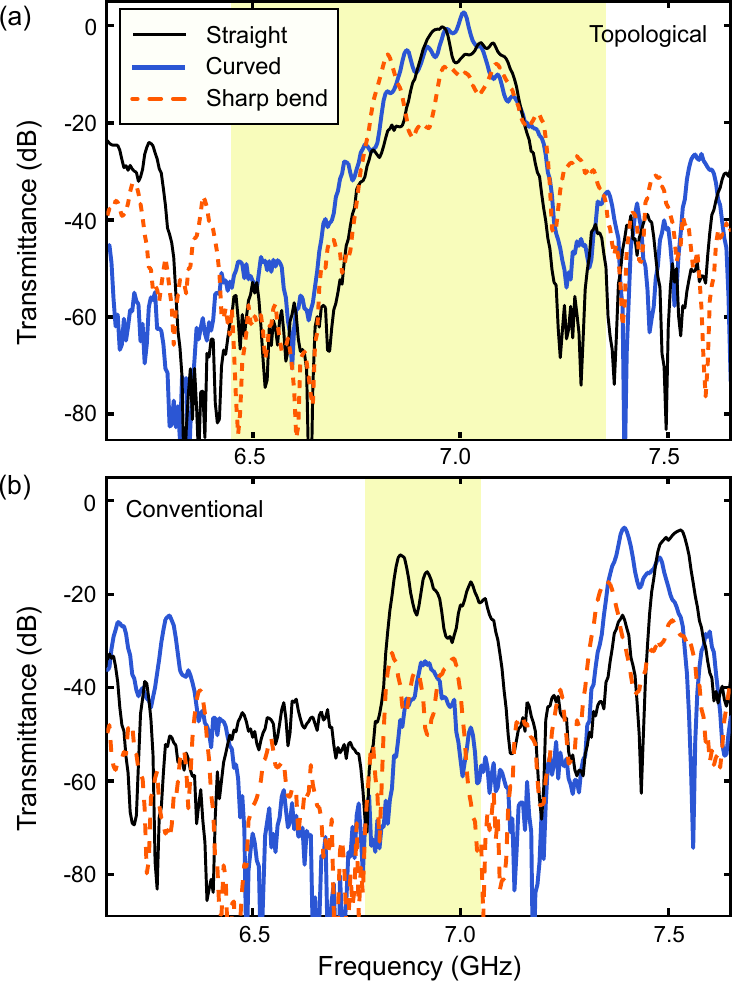}
  \caption{Experimentally measured transmittance versus frequency for topological and conventional waveguides embedded in VPCs.  For both types of waveguide, three different routes are studied: straight, curved, and sharply bent.  The configurations are the same as in the simulations of Fig.~\ref{fig:comparison}.  The transmittance is measured between two antennas located at each end of the waveguide, and all six transmittance curves are normalized to the maximum transmittance measured for the straight topological waveguide.  (a) Transmittances of topological waveguides.  Within a frequency range of around $6.8$--$7.2\,\textrm{GHz}$, all three curves are very similar.  This frequency range is narrower than the theoretically predicted band gap of $6.45$--$7.35\,\textrm{GHz}$ (marked in yellow).  (b) Transmittances of conventional waveguides.  The curved and sharply bent routes have around $20\,\textrm{dB}$ lower transmittance than the straight waveguide in the predicted operating frequency range of $6.77\,\mathrm{GHz}$--$7.05\,\textrm{GHz}$ (calculated from the equivalent semi-infinite photonic crystal waveguide, and marked in yellow).  The parameters of the VPCs are as stated in Appendix~\ref{sec:design}. }
 \label{fig:comparison_intensity}
\end{figure}

%Waveguides in VPCs can enjoy greater robustness against Anderson localization because of their links to topological valley Hall edge states.  In the crystalline case, valley Hall edge states are topologically protected against backscattering because there is one unidirectional edge state in each valley, and inter-valley scattering is suppressed when disorder length scales are sufficiently large compared to the lattice constant, as well as for 120$^\circ$ bends \cite{Ma2016}.  For VPCs with TDs, short-range order may be sufficient to provide similar robustness.  Variations in the route of the waveguide and distortions in the geometry of the surrounding lattice act like large-scale impurities, inducing negligible inter-valley scattering. Moreover, sharper waveguide bends can be implemented using local 120$^\circ$ bends, which likewise do not induce backscattering.

To investigate this empirically, we implemented several large-scale VPCs  hosting topological and non-topological waveguides.  As shown in Fig.~\ref{fig:comparison}, each waveguide forms an open arc following a straight path, a curved path, or a curved path containing a sharp bend.  Each topological waveguide runs along a disclination connecting a pentagonal and a heptagonal TD; the lattice generation procedure is similar to the one previously described (i.e., generation of an unoptimized lattice containing the desired topological defects, close packing, conversion to a dual honeycomb-like lattice, and lattice coloring).  The parameters of the VPCs are the same as before (see Appendix~\ref{sec:design}).

Figure~\ref{fig:comparison}(a)--(c) shows simulated field distributions produced by a dipole source placed at one TD.  For all three configurations, we observe a roughly uniform intensity distribution along the entire open arc of the waveguide.  The waveguides can thus be regarded as 1D Fabry-P\'erot-like optical cavities, with the topological defects serving as end mirrors (i.e., the waveguide modes experience complete inter-valley scattering and back-reflection at these points).  The uniformity of the intensity distribution indicates that there is negligible Anderson localization, as well as negligible backscattering at the sharp bend in the case of Fig.~\ref{fig:comparison}(c).

For comparison, we also implemented a set of conventional (non-topological) waveguides in similar photonic lattices.  The lattices have the same bulk parameters as before (i.e., the same pillar radii and mean inter-pillar spacings), but lack topological defects.  The waveguides do not run along disclinations, but are instead formed by the selective removal of pillars along a desired route.  In a perfectly crystalline lattice, photonic bandstructure calculations show that the pillar removal procedure creates a band of defect modes in the frequency range $6.77\,\mathrm{GHz}$--$7.05\,\textrm{GHz}$, close to the center of the bulk band gap (these are not topological modes, so they do not span the gap).  Simulations show that the defect modes transmit efficiently when the waveguide follows a straight line [Fig.~\ref{fig:comparison}(d)], but suffer from localization when the route is curved [Fig.~\ref{fig:comparison}(e)].  They are also unable to guide light efficiently around sharp corners [Fig.~\ref{fig:comparison}(f)].

We experimentally implemented these six photonic structures, and measured the transmission between the two TDs of the waveguides.  The results are shown in Fig.~\ref{fig:comparison_intensity}.  For the topological waveguides, the three routes (straight, curved, and sharply bent) all produce similar transmission characteristics at frequencies close to the center of the bulk band gap.  The operating frequency bandwidth appears to be somewhat narrower than the full width of the bulk band gap as predicted from simulations, possibly due to intrinsic losses in the ceramic material as well as input and output impedances.  For the conventional waveguides, the bent and curved routes transmit markedly less efficiently, with transmittances around $20\,\textrm{dB}$ lower than the straight conventional waveguide, and $25$--$30\,\textrm{dB}$ lower than the topological waveguides.

\begin{figure}
\centering
\includegraphics[width=\columnwidth]{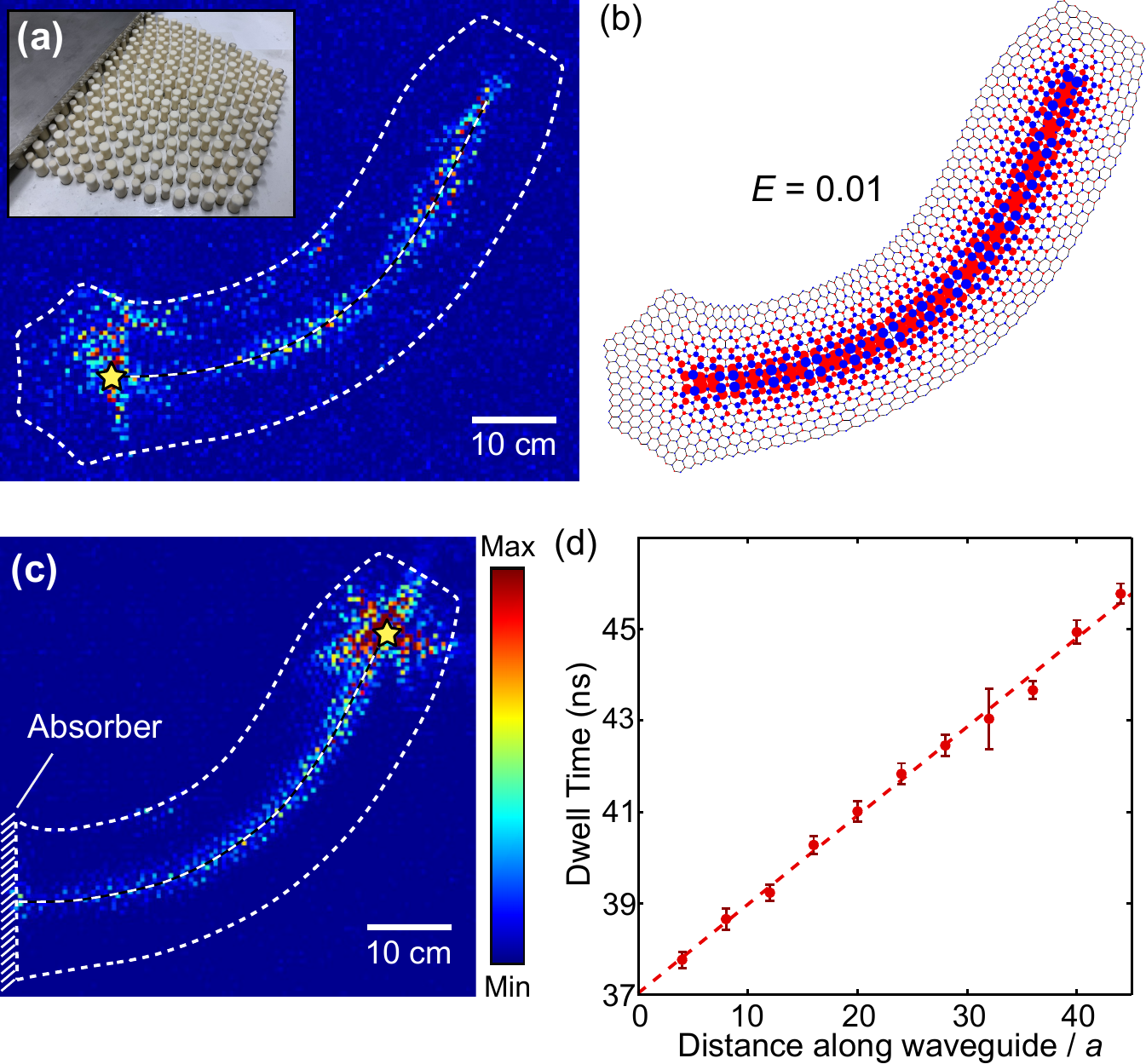}
\caption{(a) Experimentally measured distribution of the electric field intensity, $|E_z|^2$, for a topological waveguide embedded in an VPC.  The waveguide follows a curved route between a pentagonal topological defect and a heptagonal topological defect.  A $7.66\,\textrm{GHz}$ dipole source is placed near the pentagonal defect (yellow star), and the scanning resolution is $5\textrm{mm}\times5\textrm{mm}$.  Inset: photograph of the sample with the top plate partially retracted.  (b) Schematic of the lattice, overlaid with a plot of a tight-binding eigenmode with close to mid-gap energy ($E = 0.01$, with $t=1$ and $m_{A/B} = \pm 0.5$).  The radius of each plot point corresponds to the wavefunction magnitude on that site, with red and blue indicating the two sublattices.  (c) Experimentally measured electric field intensity distribution for a sample in which the waveguide terminates at an external boundary with microwave absorbing foam.  (d) Experimentally measured dwell times for the sample in (c), plotted against the distance (arc length) along the waveguide.  Each data point is the mean of 61 dwell time estimates obtained from the phase of $E_z$ over a range of frequencies centered at $7.6\,\textrm{GHz}$ with $0.002\,\textrm{GHz}$ spacing.  The error bars indicate the standard error of the mean, and the dashes show the linear least squares fit.}
\label{fig:pairs}
\end{figure}

We also performed field-mapping experiments on a curved topological waveguide [Fig.~\ref{fig:pairs}(a)--(b)].  As shown in Fig.~\ref{fig:pairs}(a), strong field intensities are observed along the entire length of the waveguide, consistent with the behavior in simulations (Fig.~\ref{fig:comparison}).  In these experiments, there is a $1\,\textrm{mm}$ air gap between the moving top plate and the pillars, which causes a small frequency shift in the band structure.  The shifted bulk band gap is estimated to be $7.36$--$8.04\,\textrm{GHz}$ (see Appendix~\ref{sec:design}).

To further probe the nature of the waveguide modes, we studied another sample in which the curved topological waveguide terminates at an external boundary lined with microwave-absorbing foam, rather than a TD.  Due to the suppressed back-reflection, this waveguide should only contain waves traveling in one direction, away from the source.  The experimental field intensity map is shown in Fig.~\ref{fig:pairs}(c).  We then measured the dwell time at different positions along the waveguide \cite{Cheng2016}.  The dwell time is defined as $d\varphi/d\omega$, where $\varphi$ is the phase of the measured field and $\omega$ is the angular frequency.  To estimate the derivative, we approximate the derivative using a finite frequency spacing $df = 0.002\,\textrm{GHz}$.  For each data point, we use a total of 61 dwell time estimates, centered around the frequency $7.6\,\textrm{GHz}$ close to the center of the gap.  The resulting mean (and standard error of the mean) are plotted in Fig.~\ref{fig:pairs}(d).  We find that the dwell time scales linearly with distance (arc length) along the curved waveguide, consistent with the expectation that the edge states propagate ballistically.

\section{Discussion}

We have demonstrated that TDs-induced disclinations in VPCs enable a novel scheme for implementing complex large-scale photonic structures, by placing TDs at the desired end-points and adjusting the disclinations to follow the desired waveguide routes.  The waveguides enjoy much greater resistance to backscattering than freeform waveguides in amorphous photonic lattices designed without topological principles \cite{Florescu2013, Man2013}.  Though originally implemented at microwave frequencies, our all-dielectric design can be straightforwardly scaled to higher frequencies, as demonstrated by the recent implementation of VPCs in the optical regime \cite{Shalaev2019}.  From a fundamental point of view, we have demonstrated a new interplay between the real-space topology of a lattice and the momentum-space topology of Bloch wavefunctions, which is different from the previously-known topological bulk-edge correspondence principle.  This phenomenon makes the presence of TDs in 2D honeycomb lattices, previously a relatively subtle effect, now easily observable.  In future work, it will be interesting to investigate TDs and disclinations in other types of topological photonic lattices, such as higher-order topological insulators \cite{Noh2018}, as well as experimentally accessing other phenomena associated with TDs such as anomalous Aharanov-Bohm effects and localized zero modes \cite{Gonzalez1993, Lammert2000, Lammert2004, Sitenko2007, Cortijo2007, Ruegg2013, deSouza2014, Jeong2008, Vozmediano2010, Yazyev2010, Kotakoski2011, wei2012}.

\appendix

\section{Effects of Topological Lattice Defects on Dirac Cone States}
\label{sec:dirac-cone-appendix}

In this Appendix, we describe the continuum limit for 2D honeycomb lattices containing topological defects, and the role of the disclinations attached to those defects.

\begin{figure}
\centering
\includegraphics[width=0.5\textwidth]{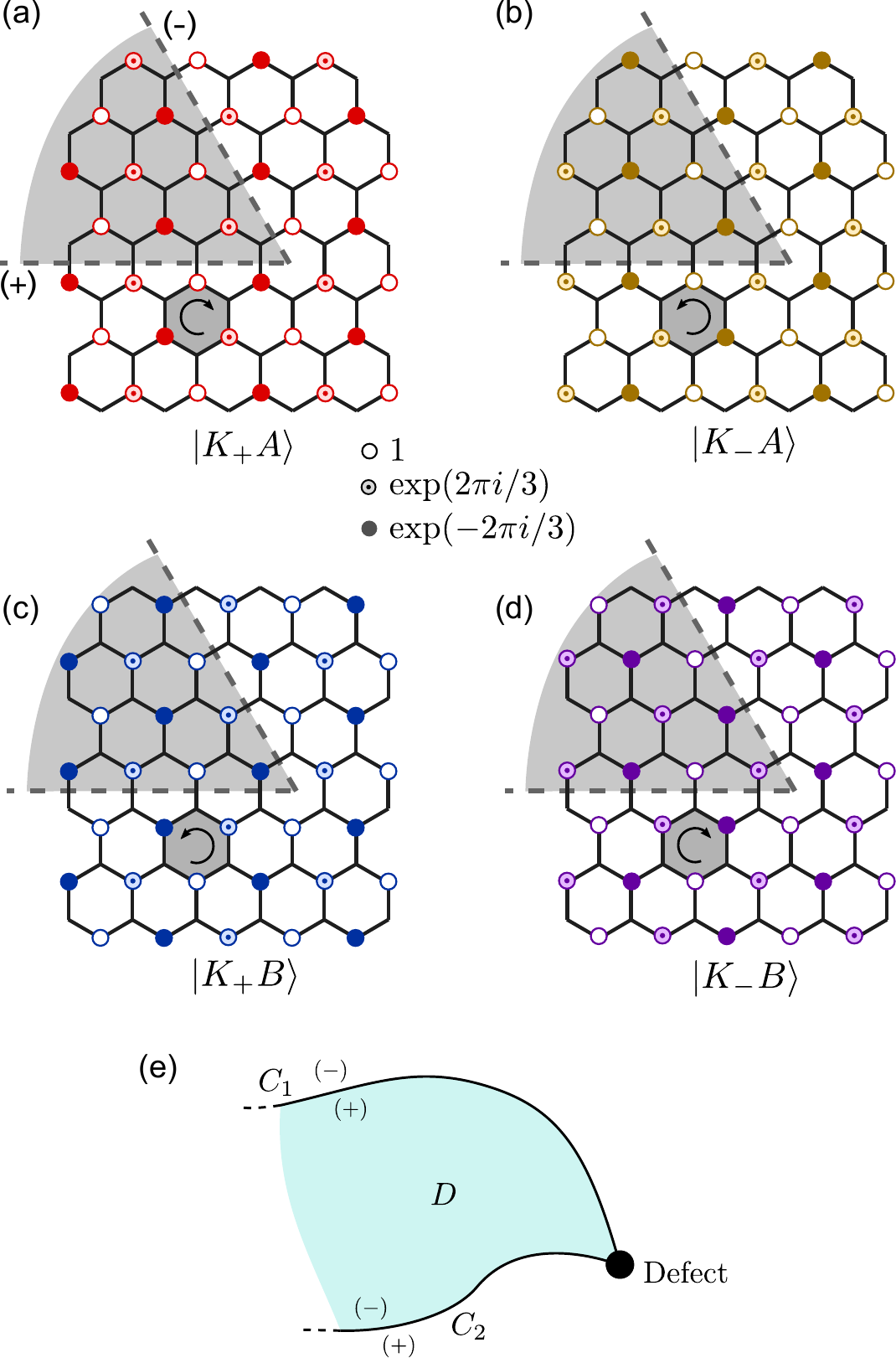}
\caption{(a)--(d) Definition of basis functions on a lattice with a pentagonal defect.  Starting from a honeycomb lattice, four Dirac point states  $|K_{\pm},A/B\rangle$ are defined by starting from a unit cell (gray  hexagon) and assigning site amplitudes $1$ (empty circles),  $\exp(2\pi i/3)$ (dotted circles), and $\exp(-2\pi i/3)$ (filled  circles) according to the given patterns, until the lattice is  covered.  Sites missing circles have zero amplitude.  The pentagonal  defect is then generated by removing a $\pi/3$ sector and  re-attaching the lattice along the seams (dashed lines).  (e) Schematic of two possible disclination strings, $C_1$ and $C_2$, bounded by domain $D$.  For $m = 0$, $C_1$ can be shifted to $C_2$ with no physical consequences.}
\label{fig:blochstates}
\end{figure}

We begin with a perfect honeycomb lattice without topological defects.  Let $|K_\pm, A/B\rangle$ denote a set of Dirac point states, where $\pm$ is the valley index and $A/B$ is the sublattice index.  As shown in Fig.~\ref{fig:blochstates}(a)--(d), these states can be defined so that the amplitudes on the lattice sites are $\{1, \eta, \eta^*\}$, where $\eta \equiv\exp(2 \pi i / 3)$, with each Dirac point state occupying a single sublattice, $A$ or $B$.  Note that the two states in each valley have opposite chirality.  The basis wavefunctions, $\Psi_{\pm A/B}(r) \equiv \langle r | K_\pm, A/B\rangle$, obey the time-reversal relation
\begin{equation}
  \Psi_{\pm A} = \Psi_{\mp A}^*,\;\;\;
  \Psi_{\pm B} = \Psi_{\mp B}^*.
  \label{timereversal}
\end{equation}
Microscopic wavefunctions can be expressed as
\begin{equation}
  \varphi(r) = \sum_{s = \pm} \sum_{\mu = A,B} \psi_{s\mu}(r) \; \Psi_{K_s\mu}(r),
  \label{envelope}
\end{equation}
where the $\psi_{s\mu}(r)$'s are slowly-varying envelope functions.  These form a four-component spinor field $\psi(r)$ that is governed by a 2D Dirac equation
$\mathcal{H}_0 \psi(r) = \epsilon\, \psi(r)$, where
\begin{equation}
  \mathcal{H}_0 = -i v \big( \tau_3 \sigma_1 \partial_1 + \sigma_2 \partial_2 \big)
  + m \sigma_3.
  \label{H0}
\end{equation}
Here $v$ is the Dirac velocity, $\tau_i$ ($\sigma_i$) denote valley (sublattice) Pauli matrices, $m$ is the sublattice detuning, and $\epsilon$ is the eigenfrequency or eigenenergy.

Next, we introduce a topological defect, e.g.~by deleting the $\pi/3$ sector marked in gray in Fig.~\ref{fig:blochstates}(a)--(d).  We can use the same pattern of site amplitudes as before to define a set of basis functions for the altered lattice \cite{Lammert2000,Lammert2004}.  Then Eq.~\eqref{envelope} still applies, but the Dirac equation governing $\psi(r)$ may be changed.  There are two distinct modifications to account for: (i) the lattice is spatially distorted in order to join up the seams of the deleted (or inserted) sector, which alters the effective Hamiltonian; (ii) the wavefunctions must satisfy some boundary condition along the disclination.

We consider these two issues in turn.  The distortion of the lattice can be modeled as a local frame rotation, which manifests in the effective Hamiltonian as follows:
\begin{align}
  \begin{aligned}
    \mathcal{H} &= -i v
    \big( \tau_3\, \sigma_1\, \partial_1'
    + \sigma_2\, \partial_2' \big)
  - m \sigma_3, \\      % Does here should change the sign of mass term? I have add a minus sign here%
  \begin{pmatrix} \partial_1' \\ \partial_2'\end{pmatrix}
    &= \begin{pmatrix}
      \cos \Theta(r) & - \sin \Theta(r) \\
      \sin \Theta(r) &  \cos \Theta(r)
    \end{pmatrix}
    \begin{pmatrix} \partial_1 \\ \partial_2\end{pmatrix},
  \end{aligned}
  \label{H_rotated}
\end{align}
where $\Theta(r)$ is a position-dependent frame rotation angle.  The lattice strain may also introduce other changes to $\mathcal{H}$, such as additional synthetic gauge fields, which we shall neglect.  The frame rotation angle is discontinous across the disclination, with
\begin{equation}
  \Theta^{(+)} - \Theta^{(-)} = - \alpha + \beta - \pi = \pm \frac{\pi}{3}
  \mod 2\pi,
  \label{theta_discontinuity}
\end{equation}
where $\pm$ applies to a pentagonal (heptagonal) defect.

Next, we determine the boundary conditions along the disclination.  An examination of Fig.~\ref{fig:blochstates}(a)--(d) indicates that
\begin{align*}
  \begin{aligned}
  \Psi_{+A}^{(-)} &= \eta^*\, \Psi_{-B}^{(+)}\,,
  \quad\;\Psi_{+B}^{(-)} = \eta \, \Psi_{-A}^{(+)}\,, \\
  \Psi_{-A}^{(-)} &= \eta\, \Psi_{+B}^{(+)}\,,
  \quad\;\;\, \Psi_{-B}^{(-)} = \eta^*\, \Psi_{+A}^{(+)},
  \end{aligned}
\end{align*}
where the $(-)$ and $(+)$ superscripts respectively indicate the clockwise and counterclockwise sides of the seams.  However, these coefficients are specific to this choice of defect position; the more general form of the disclination boundary condition is
\begin{equation}
  \begin{pmatrix}
    \Psi_{+A} \\ \Psi_{+B} \\ \Psi_{-A} \\ \Psi_{-B}
  \end{pmatrix}^{\!\!(-)} =
  \mathbf{W}^T \begin{pmatrix}
    \Psi_{+A} \\ \Psi_{+B} \\ \Psi_{-A} \\ \Psi_{-B}
  \end{pmatrix}^{\!\!(+)},
  \label{remapping1}
\end{equation}
where
\begin{equation}
  \mathbf{W} = \begin{pmatrix}
    0 & 0 & 0 & e^{i\alpha} \\
    0 & 0 & e^{i\beta} & 0 \\
    0 & e^{-i\alpha} & 0 & 0 \\
    e^{-i\beta} & 0 & 0 & 0
  \end{pmatrix},
  \label{remapping2}
\end{equation}
with $e^{i(\alpha-\beta)} = \eta (\eta^*)$ for a pentagonal (heptagonal) defect \cite{Ruegg2013}.  Note that this holds even if the disclination is not straight.  The form of $\mathbf{W}$ is constrained by the time-reversal condition \eqref{timereversal}, and by the fact that crossing the disclination swaps valley and sublattice indices, leaving the chirality unchanged.

The boundary condition for the envelope functions is
\begin{equation}
  \psi^{(+)} = \mathbf{W} \psi^{(-)},
  \label{psi_discontinuity}
\end{equation}
where $\mathbf{W}$ is the matrix defined in Eq.~\eqref{remapping2}. This can be deduced from Eqs.~\eqref{envelope} and \eqref{remapping1}, by requiring that the microscopic wavefunction $\varphi(r)$ be continuous across the disclination \cite{Gonzalez1993,Lammert2000,Lammert2004}.

Note that upon defining
\begin{equation*}
  \tilde\psi(r) = \mathbf{W}^\dagger \psi(r),
\end{equation*}
the transformed field $\tilde{\psi}$ obeys Eq.~\eqref{H_rotated} with
\begin{align}
  \begin{aligned}
    \epsilon &\rightarrow \epsilon \\
    m &\rightarrow -m \\
    \Theta(r) &\rightarrow \Theta(r) + \alpha - \beta + \pi.
  \end{aligned}
  \label{theta_rotation}
\end{align}

%% \begin{figure}
%% \centering
%% \includegraphics[width=\columnwidth]{fig_strings}
%% \caption{Two possible disclination strings, $C_1$ and $C_2$, bounded
%%   by domain $D$.  For $m = 0$, $C_1$ can be shifted to $C_2$ with no
%%   physical consequences.}
%% \label{fig:strings}
%% \end{figure}

As discussed in Section~\ref{sec:structure}, the position of the disclination has no physical effects for $m = 0$.  To see this, consider the scenario shown in Fig.~\ref{fig:blochstates}(d).  Let $\psi(r)$ be a solution in which Eqs.~\eqref{theta_discontinuity} and \eqref{psi_discontinuity} hold along a disclination $C_1$.  If we want to move the disclination to $C_2$, let $D$ be the area bounded between $C_1$ and $C_2$, and let
\begin{equation}
  \tilde{\psi}(r) =
  \begin{cases}
    \mathbf{W}^\dagger \psi(r), & r \in D \\
    \psi(r), & r \notin D.
  \end{cases}
\end{equation}
Then,
\begin{align*}
  \begin{aligned}
    \mathrm{Along} \;C_1: &\;\;
    \tilde{\psi}^{(+)} =
    \mathbf{W}^\dagger \psi^{(+)} =
    \mathbf{W}^\dagger \mathbf{W} \psi^{(-)} = \tilde{\psi}^{(-)} \\
    \mathrm{Along} \;C_2: &\;\;
    \tilde{\psi}^{(+)} = \psi^{(+)} = \psi^{(-)} = \mathbf{W} \tilde{\psi}^{(-)}.
  \end{aligned}
\end{align*}
Hence, the condition \eqref{psi_discontinuity} has shifted from $C_1$ to $C_2$.  Likewise, we can use Eq.~\eqref{theta_rotation} to show that $\Theta$ now obeys Eq.~\eqref{theta_discontinuity} along $C_2$ and is continuous along $C_1$.

The model also points to the existence of topological edge states along the disclination when $m \ne 0$.  Near the disclination, and sufficiently far from the topological defect, we can apply the transformation \eqref{remapping1} to one side of the string; this switches the sign of $m$ on that side, while making the wavefunctions and frame rotation angles continuous along the disclination.  This is locally equivalent to having two decoupled valleys with opposite signs of $m$ on opposite sides of the disclination.

\section{Design of the Photonic Lattice}
\label{sec:design}

\begin{figure}
\centering
\includegraphics[width=0.5\textwidth]{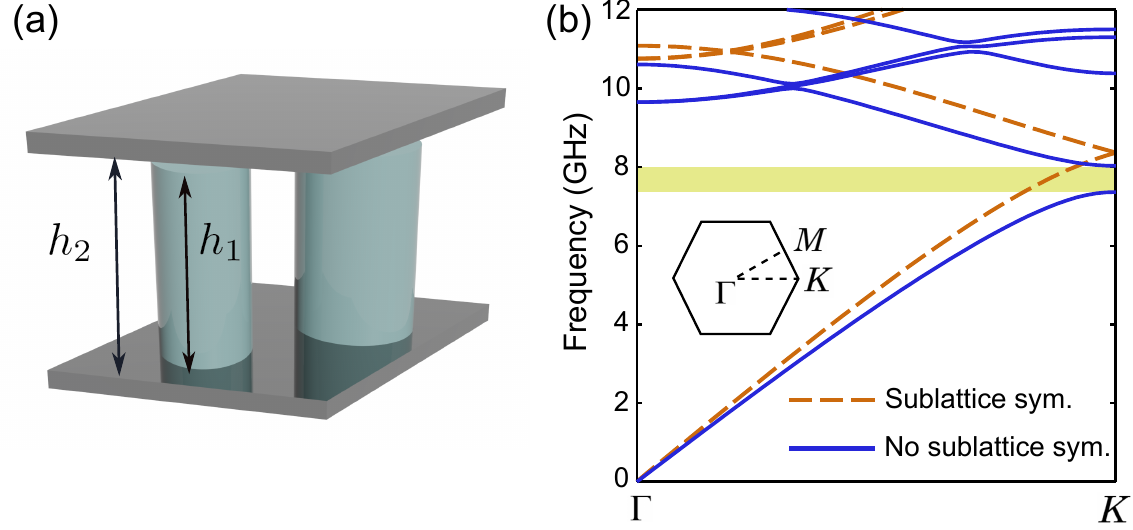}
\caption{(a) Schematic of a pair of neighboring pillars in the photonic lattice, with radius $r_{a} = 2.35\,\textrm{mm}$ and $r_{b} = 3.15\,\textrm{mm}$, height $h_{1} = 10\,\textrm{mm}$, and refractive index $n = 2.915$.  The spacing between the top and bottom metal plates is $h_{2}$.  (b) Bandstructure of the corresponding photonic crystal, calculated in 3D using $h_{2} = 11\,\textrm{mm}$ and lattice constant $a = 13.7\,\textrm{mm}$.  Two cases are plotted: all pillars having radius $r_a$ (orange dashes), and the pillars having radii $r_a$ and $r_b$ on the two sublattices (blue curves).  In the latter case, there is a photonic band gap at $7.36$--$8.04\,\textrm{GHz}$ (marked in yellow).}
\label{fig:dispersion}
\end{figure}

The valley hall lattices are implemented experimentally using ceramic pillars of refractive index $n = 2.915$, arranged in the space between parallel metal plates [Fig.~\ref{fig:dispersion}(a)].  Since the experiment occurs in the microwave regime, the top and bottom plates act as perfect electrical conductors.  The pillars are placed in a honeycomb-like  lattice, have have radii $r_{a} = 2.35\,\textrm{mm}$ and $r_{b} = 3.15\,\textrm{mm}$ for the two sublattices.  As the lattice is amorphous, there is a distribution of pillar separations; the lattice is scaled so the mean next nearest neighbor distance (roughly speaking, the lattice constant of the equivalent honeycomb lattice) is $a = 13.7\,\textrm{mm}$.  All pillars have height $h_{1} = 10\,\textrm{mm}$.

For the static transmission measurements of Fig.~\ref{fig:comparison_intensity}, we set $h_1 = h_2 = 10\,\textrm{mm}$ (i.e., the plates directly touch the pillars).  Calculations on the equivalent 3D photonic crystal (i.e., an infinite perfectly crystalline lattice with the same structural parameters) show that the band structure is almost identical to that of a 2D lattice (like in the simulations of Fig.~\ref{fig:comparison}), with band gap at $6.45$--$7.35\,\textrm{GHz}$.

For the field-mapping experiments (Figs.~\ref{fig:experiment_pentagon} and \ref{fig:pairs}), we take $h_2 = 11\,\textrm{mm}$ (i.e., there is a $1\,\textrm{mm}$ air gap to avoid direct mechanical contact between the moving top plate and the pillars).  The air gap induces a slight frequency shift relative to the 2D band structure.  Fig.~\ref{fig:dispersion}(b) plots the in-plane band structure of the equivalent crystal lattice with lattice constant $a = 13.7\,\textrm{mm}$.  If the pillars on both sublattices have radius $r_{a}$, there are two Dirac points at $8.36\,\textrm{GHz}$ at the corners of the hexagonal Brillouin zone.  If the pillars on the two sublattices have radii $r_a$ and $r_b$, the band gap occurs at $7.36$--$8.04\,\textrm{GHz}$.

\begin{figure}[b]
\centering
\includegraphics[width=0.5\textwidth]{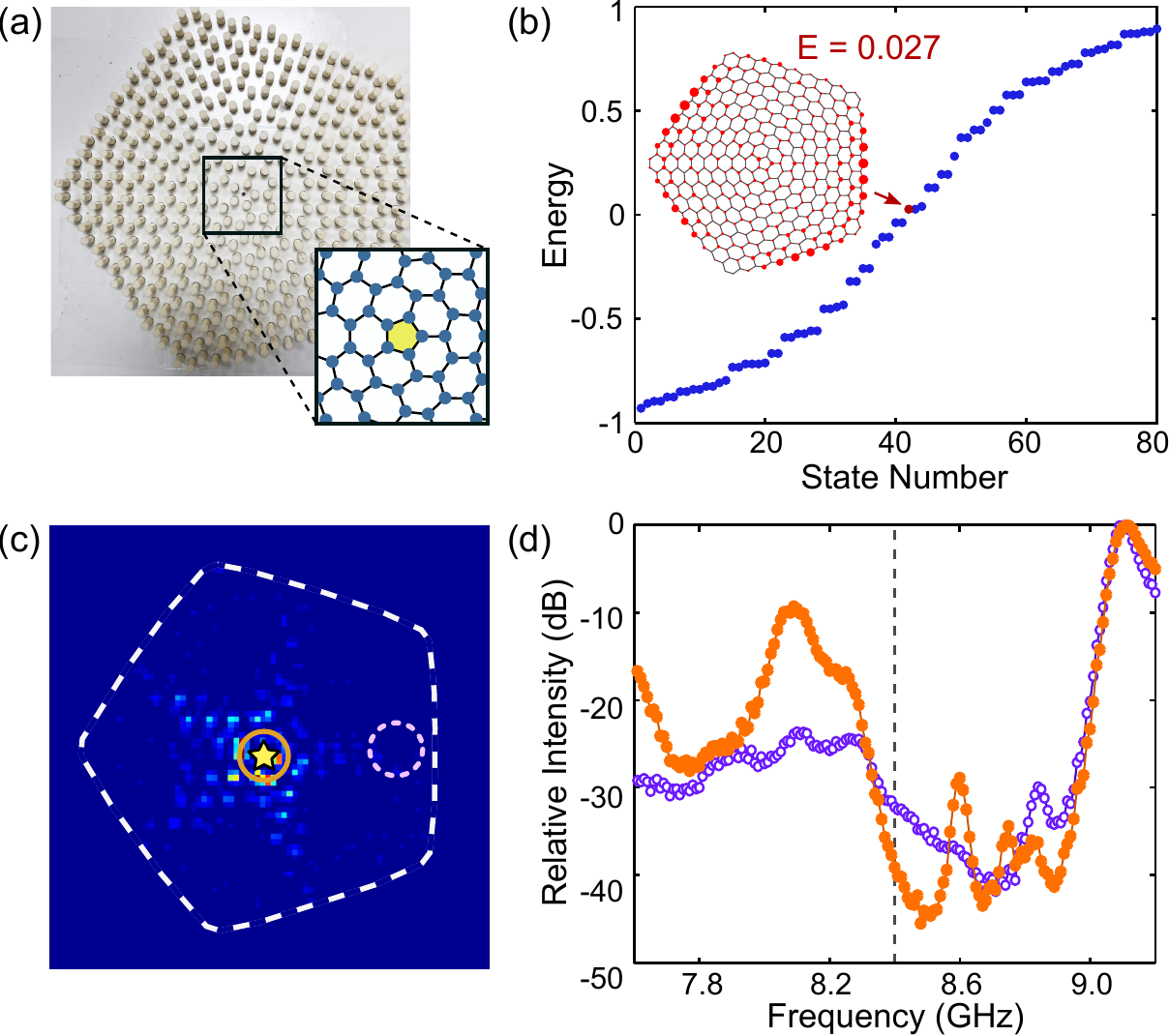}
\caption{(a) Photograph of the amorphous photonic graphene sample. Inset: schematic of the pentagonal topological defect at the center of the sample.  (b) Energy spectrum of a tight-binding model based on the given lattice, with nearest neighbor hopping $t = 1$ and on-site mass $m = 0$.  There is no eigenstate localized at the topological defect.  Inset: mode intensity for the eigenstate with energy closest to zero ($E = 0.027$).  (c) Experimentally measured distribution of the field intensity, $|E_z|^2$, with a $8.39\,\textrm{GHz}$ dipole source placed at the center of the sample (yellow star).  (d) Frequency dependence of the field intensities averaged within two regions of the lattice, surrounding the topological defect (orange dots) and away from the defect (purple dots).  These regions are respectively indicated by the solid and dashed circles in (c).  Vertical dashes mark the Dirac frequency of the photonic crystal. }
\label{fig:dirac}
\end{figure}

\section{TDs in Photonic Graphene}
\label{sec:graphene}

We fabricated an experimental sample based on an ``photonic graphene'' with  a pentagonal TD in the center [Fig.~\ref{fig:dirac}(a)], i.e all pillars having equal radius $r_a$.  To help understand the qualitative implications of this lattice geometry, Fig.~\ref{fig:dirac}(b) plots the energy spectrum of a tight-binding model based on this lattice, with nearest neighbor hopping $t = 1$ and on-site mass $m = 0$.  As expected, there is no band gap.  We examined the individual eigenstates and found none that are strongly localized at the TD; the intensity distribution of a typical eigenstate (with energy $E = -0.065$) is shown in the inset to Fig.~\ref{fig:dirac}(b).  This is consistent with previous theoretical analyses of topological defects in graphene \cite{Lammert2000, Lammert2004, Ruegg2013}, which showed that such defects act upon Dirac cone states like a gauge field rather than a confining potential.

Fig.~\ref{fig:dirac}(c) plots the measured field intensity map with a dipole source (oriented perpendicular to the plane) placed at the TD.  The source frequency is $8.39\,\textrm{GHz}$, close to the predicted Dirac frequency of $8.36\,\textrm{GHz}$ [see Fig.~\ref{fig:dispersion}(b)].  Unlike the VPCs shown in Fig.~\ref{fig:experiment_pentagon} in manuscript, no waveguiding is observed due to the lack of a disclination.  The frequency dependence of the field intensity, plotted in Fig.~\ref{fig:dirac}(d), shows no apparent resonance associated with the topological defect.

R\"uegg and Lin have found that if T is broken, topological defects in such lattices can generate localized bound states \cite{Ruegg2013}.  Interestingly, such bound states are tied to the Chern number of the surrounding lattice and are robust against perturbations.  In future work, photonic lattices could be used to to realize this theoretical prediction, either using magneto-optic materials to directly break T, or using three-dimensional structures in which an axial direction serves as an effective time \cite{Ozawa2019}.

\bibliography{citepaper.bib}

\begin{thebibliography}{52}%
\makeatletter
\providecommand \@ifxundefined [1]{%
 \@ifx{#1\undefined}
}%
\providecommand \@ifnum [1]{%
 \ifnum #1\expandafter \@firstoftwo
 \else \expandafter \@secondoftwo
 \fi
}%
\providecommand \@ifx [1]{%
 \ifx #1\expandafter \@firstoftwo
 \else \expandafter \@secondoftwo
 \fi
}%
\providecommand \natexlab [1]{#1}%
\providecommand \enquote  [1]{``#1''}%
\providecommand \bibnamefont  [1]{#1}%
\providecommand \bibfnamefont [1]{#1}%
\providecommand \citenamefont [1]{#1}%
\providecommand \href@noop [0]{\@secondoftwo}%
\providecommand \href [0]{\begingroup \@sanitize@url \@href}%
\providecommand \@href[1]{\@@startlink{#1}\@@href}%
\providecommand \@@href[1]{\endgroup#1\@@endlink}%
\providecommand \@sanitize@url [0]{\catcode `\\12\catcode `\$12\catcode
  `\&12\catcode `\#12\catcode `\^12\catcode `\_12\catcode `\%12\relax}%
\providecommand \@@startlink[1]{}%
\providecommand \@@endlink[0]{}%
\providecommand \url  [0]{\begingroup\@sanitize@url \@url }%
\providecommand \@url [1]{\endgroup\@href {#1}{\urlprefix }}%
\providecommand \urlprefix  [0]{URL }%
\providecommand \Eprint [0]{\href }%
\providecommand \doibase [0]{http://dx.doi.org/}%
\providecommand \selectlanguage [0]{\@gobble}%
\providecommand \bibinfo  [0]{\@secondoftwo}%
\providecommand \bibfield  [0]{\@secondoftwo}%
\providecommand \translation [1]{[#1]}%
\providecommand \BibitemOpen [0]{}%
\providecommand \bibitemStop [0]{}%
\providecommand \bibitemNoStop [0]{.\EOS\space}%
\providecommand \EOS [0]{\spacefactor3000\relax}%
\providecommand \BibitemShut  [1]{\csname bibitem#1\endcsname}%
\let\auto@bib@innerbib\@empty
%</preamble>
\bibitem [{\citenamefont {Ozawa}\ \emph {et~al.}(2019)\citenamefont {Ozawa},
  \citenamefont {Price}, \citenamefont {Amo}, \citenamefont {Goldman},
  \citenamefont {Hafezi}, \citenamefont {Lu}, \citenamefont {Rechtsman},
  \citenamefont {Schuster}, \citenamefont {Simon}, \citenamefont {Zilberberg},\
  and\ \citenamefont {Carusotto}}]{Ozawa2019}%
  \BibitemOpen
  \bibfield  {author} {\bibinfo {author} {\bibfnamefont {T.}~\bibnamefont
  {Ozawa}}, \bibinfo {author} {\bibfnamefont {H.~M.}\ \bibnamefont {Price}},
  \bibinfo {author} {\bibfnamefont {A.}~\bibnamefont {Amo}}, \bibinfo {author}
  {\bibfnamefont {N.}~\bibnamefont {Goldman}}, \bibinfo {author} {\bibfnamefont
  {M.}~\bibnamefont {Hafezi}}, \bibinfo {author} {\bibfnamefont
  {L.}~\bibnamefont {Lu}}, \bibinfo {author} {\bibfnamefont {M.~C.}\
  \bibnamefont {Rechtsman}}, \bibinfo {author} {\bibfnamefont {D.}~\bibnamefont
  {Schuster}}, \bibinfo {author} {\bibfnamefont {J.}~\bibnamefont {Simon}},
  \bibinfo {author} {\bibfnamefont {O.}~\bibnamefont {Zilberberg}}, \ and\
  \bibinfo {author} {\bibfnamefont {I.}~\bibnamefont {Carusotto}},\ }\href
  {\doibase 10.1103/RevModPhys.91.015006} {\bibfield  {journal} {\bibinfo
  {journal} {Rev. Mod. Phys.}\ }\textbf {\bibinfo {volume} {91}},\ \bibinfo
  {pages} {015006} (\bibinfo {year} {2019})}\BibitemShut {NoStop}%
\bibitem [{\citenamefont {Bansil}\ \emph {et~al.}(2016)\citenamefont {Bansil},
  \citenamefont {Lin},\ and\ \citenamefont {Das}}]{Bansil2016}%
  \BibitemOpen
  \bibfield  {author} {\bibinfo {author} {\bibfnamefont {A.}~\bibnamefont
  {Bansil}}, \bibinfo {author} {\bibfnamefont {H.}~\bibnamefont {Lin}}, \ and\
  \bibinfo {author} {\bibfnamefont {T.}~\bibnamefont {Das}},\ }\href {\doibase
  10.1103/RevModPhys.88.021004} {\bibfield  {journal} {\bibinfo  {journal}
  {Rev. Mod. Phys.}\ }\textbf {\bibinfo {volume} {88}},\ \bibinfo {pages}
  {021004} (\bibinfo {year} {2016})}\BibitemShut {NoStop}%
\bibitem [{\citenamefont {Hafezi}\ \emph {et~al.}(2011)\citenamefont {Hafezi},
  \citenamefont {Demler}, \citenamefont {Lukin},\ and\ \citenamefont
  {Taylor}}]{Hafezi2011}%
  \BibitemOpen
  \bibfield  {author} {\bibinfo {author} {\bibfnamefont {M.}~\bibnamefont
  {Hafezi}}, \bibinfo {author} {\bibfnamefont {E.~A.}\ \bibnamefont {Demler}},
  \bibinfo {author} {\bibfnamefont {M.~D.}\ \bibnamefont {Lukin}}, \ and\
  \bibinfo {author} {\bibfnamefont {J.~M.}\ \bibnamefont {Taylor}},\ }\href
  {\doibase 10.1038/nphys2063} {\bibfield  {journal} {\bibinfo  {journal} {Nat.
  Phys.}\ }\textbf {\bibinfo {volume} {7}},\ \bibinfo {pages} {907} (\bibinfo
  {year} {2011})}\BibitemShut {NoStop}%
\bibitem [{\citenamefont {Wang}\ \emph {et~al.}(2008)\citenamefont {Wang},
  \citenamefont {Chong}, \citenamefont {Joannopoulos},\ and\ \citenamefont
  {Solja\ifmmode \check{c}\else \v{c}\fi{}i\ifmmode~\acute{c}\else
  \'{c}\fi{}}}]{Wang2008}%
  \BibitemOpen
  \bibfield  {author} {\bibinfo {author} {\bibfnamefont {Z.}~\bibnamefont
  {Wang}}, \bibinfo {author} {\bibfnamefont {Y.~D.}\ \bibnamefont {Chong}},
  \bibinfo {author} {\bibfnamefont {J.~D.}\ \bibnamefont {Joannopoulos}}, \
  and\ \bibinfo {author} {\bibfnamefont {M.}~\bibnamefont {Solja\ifmmode
  \check{c}\else \v{c}\fi{}i\ifmmode~\acute{c}\else \'{c}\fi{}}},\ }\href
  {\doibase 10.1103/PhysRevLett.100.013905} {\bibfield  {journal} {\bibinfo
  {journal} {Phys. Rev. Lett.}\ }\textbf {\bibinfo {volume} {100}},\ \bibinfo
  {pages} {013905} (\bibinfo {year} {2008})}\BibitemShut {NoStop}%
\bibitem [{\citenamefont {Wang}\ \emph {et~al.}(2009)\citenamefont {Wang},
  \citenamefont {Chong}, \citenamefont {Joannopoulos},\ and\ \citenamefont
  {Soljačić}}]{Wang2009}%
  \BibitemOpen
  \bibfield  {author} {\bibinfo {author} {\bibfnamefont {Z.}~\bibnamefont
  {Wang}}, \bibinfo {author} {\bibfnamefont {Y.}~\bibnamefont {Chong}},
  \bibinfo {author} {\bibfnamefont {J.~D.}\ \bibnamefont {Joannopoulos}}, \
  and\ \bibinfo {author} {\bibfnamefont {M.}~\bibnamefont {Soljačić}},\
  }\href {\doibase 10.1038/nature08293} {\bibfield  {journal} {\bibinfo
  {journal} {Nature}\ }\textbf {\bibinfo {volume} {461}},\ \bibinfo {pages}
  {772} (\bibinfo {year} {2009})}\BibitemShut {NoStop}%
\bibitem [{\citenamefont {Dong}\ \emph {et~al.}(2017)\citenamefont {Dong},
  \citenamefont {Chen}, \citenamefont {Zhu}, \citenamefont {Wang},\ and\
  \citenamefont {Zhang}}]{Dong2017}%
  \BibitemOpen
  \bibfield  {author} {\bibinfo {author} {\bibfnamefont {J.-W.}\ \bibnamefont
  {Dong}}, \bibinfo {author} {\bibfnamefont {X.-D.}\ \bibnamefont {Chen}},
  \bibinfo {author} {\bibfnamefont {H.}~\bibnamefont {Zhu}}, \bibinfo {author}
  {\bibfnamefont {Y.}~\bibnamefont {Wang}}, \ and\ \bibinfo {author}
  {\bibfnamefont {X.}~\bibnamefont {Zhang}},\ }\href {\doibase
  10.1038/nmat4807} {\bibfield  {journal} {\bibinfo  {journal} {Nat. Mater.}\
  }\textbf {\bibinfo {volume} {16}},\ \bibinfo {pages} {298} (\bibinfo {year}
  {2017})}\BibitemShut {NoStop}%
\bibitem [{\citenamefont {Shalaev}\ \emph {et~al.}(2019)\citenamefont
  {Shalaev}, \citenamefont {Walasik}, \citenamefont {Tsukernik}, \citenamefont
  {Xu},\ and\ \citenamefont {Litchinitser}}]{Shalaev2019}%
  \BibitemOpen
  \bibfield  {author} {\bibinfo {author} {\bibfnamefont {M.~I.}\ \bibnamefont
  {Shalaev}}, \bibinfo {author} {\bibfnamefont {W.}~\bibnamefont {Walasik}},
  \bibinfo {author} {\bibfnamefont {A.}~\bibnamefont {Tsukernik}}, \bibinfo
  {author} {\bibfnamefont {Y.}~\bibnamefont {Xu}}, \ and\ \bibinfo {author}
  {\bibfnamefont {N.~M.}\ \bibnamefont {Litchinitser}},\ }\href {\doibase
  10.1038/s41565-018-0297-6} {\bibfield  {journal} {\bibinfo  {journal} {Nat.
  Nanotech.}\ }\textbf {\bibinfo {volume} {14}},\ \bibinfo {pages} {31}
  (\bibinfo {year} {2019})}\BibitemShut {NoStop}%
\bibitem [{\citenamefont {Hadad}\ \emph {et~al.}(2018)\citenamefont {Hadad},
  \citenamefont {Soric}, \citenamefont {Khanikaev},\ and\ \citenamefont
  {Alù}}]{Hadad2018}%
  \BibitemOpen
  \bibfield  {author} {\bibinfo {author} {\bibfnamefont {Y.}~\bibnamefont
  {Hadad}}, \bibinfo {author} {\bibfnamefont {J.~C.}\ \bibnamefont {Soric}},
  \bibinfo {author} {\bibfnamefont {A.~B.}\ \bibnamefont {Khanikaev}}, \ and\
  \bibinfo {author} {\bibfnamefont {A.}~\bibnamefont {Alù}},\ }\href {\doibase
  10.1038/s41928-018-0042-z} {\bibfield  {journal} {\bibinfo  {journal} {Nat.
  Electron.}\ }\textbf {\bibinfo {volume} {1}},\ \bibinfo {pages} {178}
  (\bibinfo {year} {2018})}\BibitemShut {NoStop}%
\bibitem [{\citenamefont {Wang}\ \emph {et~al.}(2019)\citenamefont {Wang},
  \citenamefont {Lang}, \citenamefont {Lee}, \citenamefont {Zhang},\ and\
  \citenamefont {Chong}}]{wang2019}%
  \BibitemOpen
  \bibfield  {author} {\bibinfo {author} {\bibfnamefont {Y.}~\bibnamefont
  {Wang}}, \bibinfo {author} {\bibfnamefont {L.-J.}\ \bibnamefont {Lang}},
  \bibinfo {author} {\bibfnamefont {C.~H.}\ \bibnamefont {Lee}}, \bibinfo
  {author} {\bibfnamefont {B.}~\bibnamefont {Zhang}}, \ and\ \bibinfo {author}
  {\bibfnamefont {Y.~D.}\ \bibnamefont {Chong}},\ }\href {\doibase
  10.1038/s41467-019-08966-9} {\bibfield  {journal} {\bibinfo  {journal} {Nat.
  Commun.}\ }\textbf {\bibinfo {volume} {10}},\ \bibinfo {pages} {1102}
  (\bibinfo {year} {2019})}\BibitemShut {NoStop}%
\bibitem [{\citenamefont {St-Jean}\ \emph {et~al.}(2017)\citenamefont
  {St-Jean}, \citenamefont {Goblot}, \citenamefont {Galopin}, \citenamefont
  {Lemaître}, \citenamefont {Ozawa}, \citenamefont {Le~Gratiet}, \citenamefont
  {Sagnes}, \citenamefont {Bloch},\ and\ \citenamefont {Amo}}]{StJean2017}%
  \BibitemOpen
  \bibfield  {author} {\bibinfo {author} {\bibfnamefont {P.}~\bibnamefont
  {St-Jean}}, \bibinfo {author} {\bibfnamefont {V.}~\bibnamefont {Goblot}},
  \bibinfo {author} {\bibfnamefont {E.}~\bibnamefont {Galopin}}, \bibinfo
  {author} {\bibfnamefont {A.}~\bibnamefont {Lemaître}}, \bibinfo {author}
  {\bibfnamefont {T.}~\bibnamefont {Ozawa}}, \bibinfo {author} {\bibfnamefont
  {L.}~\bibnamefont {Le~Gratiet}}, \bibinfo {author} {\bibfnamefont
  {I.}~\bibnamefont {Sagnes}}, \bibinfo {author} {\bibfnamefont
  {J.}~\bibnamefont {Bloch}}, \ and\ \bibinfo {author} {\bibfnamefont
  {A.}~\bibnamefont {Amo}},\ }\href {\doibase 10.1038/s41566-017-0006-2}
  {\bibfield  {journal} {\bibinfo  {journal} {Nat. Photon}\ }\textbf {\bibinfo
  {volume} {11}},\ \bibinfo {pages} {651} (\bibinfo {year} {2017})}\BibitemShut
  {NoStop}%
\bibitem [{\citenamefont {Zhao}\ \emph {et~al.}(2018)\citenamefont {Zhao},
  \citenamefont {Miao}, \citenamefont {Teimourpour}, \citenamefont {Malzard},
  \citenamefont {El-Ganainy}, \citenamefont {Schomerus},\ and\ \citenamefont
  {Feng}}]{Zhao_Feng2018}%
  \BibitemOpen
  \bibfield  {author} {\bibinfo {author} {\bibfnamefont {H.}~\bibnamefont
  {Zhao}}, \bibinfo {author} {\bibfnamefont {P.}~\bibnamefont {Miao}}, \bibinfo
  {author} {\bibfnamefont {M.~H.}\ \bibnamefont {Teimourpour}}, \bibinfo
  {author} {\bibfnamefont {S.}~\bibnamefont {Malzard}}, \bibinfo {author}
  {\bibfnamefont {R.}~\bibnamefont {El-Ganainy}}, \bibinfo {author}
  {\bibfnamefont {H.}~\bibnamefont {Schomerus}}, \ and\ \bibinfo {author}
  {\bibfnamefont {L.}~\bibnamefont {Feng}},\ }\href {\doibase
  10.1038/s41467-018-03434-2} {\bibfield  {journal} {\bibinfo  {journal} {Nat.
  Commun.}\ }\textbf {\bibinfo {volume} {9}},\ \bibinfo {pages} {981} (\bibinfo
  {year} {2018})}\BibitemShut {NoStop}%
\bibitem [{\citenamefont {Parto}\ \emph {et~al.}(2018)\citenamefont {Parto},
  \citenamefont {Wittek}, \citenamefont {Hodaei}, \citenamefont {Harari},
  \citenamefont {Bandres}, \citenamefont {Ren}, \citenamefont {Rechtsman},
  \citenamefont {Segev}, \citenamefont {Christodoulides},\ and\ \citenamefont
  {Khajavikhan}}]{Parto2018}%
  \BibitemOpen
  \bibfield  {author} {\bibinfo {author} {\bibfnamefont {M.}~\bibnamefont
  {Parto}}, \bibinfo {author} {\bibfnamefont {S.}~\bibnamefont {Wittek}},
  \bibinfo {author} {\bibfnamefont {H.}~\bibnamefont {Hodaei}}, \bibinfo
  {author} {\bibfnamefont {G.}~\bibnamefont {Harari}}, \bibinfo {author}
  {\bibfnamefont {M.~A.}\ \bibnamefont {Bandres}}, \bibinfo {author}
  {\bibfnamefont {J.}~\bibnamefont {Ren}}, \bibinfo {author} {\bibfnamefont
  {M.~C.}\ \bibnamefont {Rechtsman}}, \bibinfo {author} {\bibfnamefont
  {M.}~\bibnamefont {Segev}}, \bibinfo {author} {\bibfnamefont {D.~N.}\
  \bibnamefont {Christodoulides}}, \ and\ \bibinfo {author} {\bibfnamefont
  {M.}~\bibnamefont {Khajavikhan}},\ }\href {\doibase
  10.1103/PhysRevLett.120.113901} {\bibfield  {journal} {\bibinfo  {journal}
  {Phys. Rev. Lett.}\ }\textbf {\bibinfo {volume} {120}},\ \bibinfo {pages}
  {113901} (\bibinfo {year} {2018})}\BibitemShut {NoStop}%
\bibitem [{\citenamefont {Ota}\ \emph {et~al.}(2018)\citenamefont {Ota},
  \citenamefont {Katsumi}, \citenamefont {Watanabe}, \citenamefont {Iwamoto},\
  and\ \citenamefont {Arakawa}}]{Ota2018}%
  \BibitemOpen
  \bibfield  {author} {\bibinfo {author} {\bibfnamefont {Y.}~\bibnamefont
  {Ota}}, \bibinfo {author} {\bibfnamefont {R.}~\bibnamefont {Katsumi}},
  \bibinfo {author} {\bibfnamefont {K.}~\bibnamefont {Watanabe}}, \bibinfo
  {author} {\bibfnamefont {S.}~\bibnamefont {Iwamoto}}, \ and\ \bibinfo
  {author} {\bibfnamefont {Y.}~\bibnamefont {Arakawa}},\ }\href {\doibase
  10.1038/s42005-018-0083-7} {\bibfield  {journal} {\bibinfo  {journal}
  {Commun. Phys.}\ }\textbf {\bibinfo {volume} {1}},\ \bibinfo {pages} {86}
  (\bibinfo {year} {2018})}\BibitemShut {NoStop}%
\bibitem [{\citenamefont {Bandres}\ \emph {et~al.}(2018)\citenamefont
  {Bandres}, \citenamefont {Wittek}, \citenamefont {Harari}, \citenamefont
  {Parto}, \citenamefont {Ren}, \citenamefont {Segev}, \citenamefont
  {Christodoulides},\ and\ \citenamefont {Khajavikhan}}]{Bandres2018}%
  \BibitemOpen
  \bibfield  {author} {\bibinfo {author} {\bibfnamefont {M.~A.}\ \bibnamefont
  {Bandres}}, \bibinfo {author} {\bibfnamefont {S.}~\bibnamefont {Wittek}},
  \bibinfo {author} {\bibfnamefont {G.}~\bibnamefont {Harari}}, \bibinfo
  {author} {\bibfnamefont {M.}~\bibnamefont {Parto}}, \bibinfo {author}
  {\bibfnamefont {J.}~\bibnamefont {Ren}}, \bibinfo {author} {\bibfnamefont
  {M.}~\bibnamefont {Segev}}, \bibinfo {author} {\bibfnamefont {D.~N.}\
  \bibnamefont {Christodoulides}}, \ and\ \bibinfo {author} {\bibfnamefont
  {M.}~\bibnamefont {Khajavikhan}},\ }\href {\doibase 10.1126/science.aar4005}
  {\bibfield  {journal} {\bibinfo  {journal} {Science}\ }\textbf {\bibinfo
  {volume} {359}},\ \bibinfo {pages} {4005} (\bibinfo {year}
  {2018})}\BibitemShut {NoStop}%
\bibitem [{\citenamefont {Harari}\ \emph {et~al.}(2018)\citenamefont {Harari},
  \citenamefont {Bandres}, \citenamefont {Lumer}, \citenamefont {Rechtsman},
  \citenamefont {Chong}, \citenamefont {Khajavikhan}, \citenamefont
  {Christodoulides},\ and\ \citenamefont {Segev}}]{Harari2018}%
  \BibitemOpen
  \bibfield  {author} {\bibinfo {author} {\bibfnamefont {G.}~\bibnamefont
  {Harari}}, \bibinfo {author} {\bibfnamefont {M.~A.}\ \bibnamefont {Bandres}},
  \bibinfo {author} {\bibfnamefont {Y.}~\bibnamefont {Lumer}}, \bibinfo
  {author} {\bibfnamefont {M.~C.}\ \bibnamefont {Rechtsman}}, \bibinfo {author}
  {\bibfnamefont {Y.~D.}\ \bibnamefont {Chong}}, \bibinfo {author}
  {\bibfnamefont {M.}~\bibnamefont {Khajavikhan}}, \bibinfo {author}
  {\bibfnamefont {D.~N.}\ \bibnamefont {Christodoulides}}, \ and\ \bibinfo
  {author} {\bibfnamefont {M.}~\bibnamefont {Segev}},\ }\href {\doibase
  10.1126/science.aar4003} {\bibfield  {journal} {\bibinfo  {journal}
  {Science}\ }\textbf {\bibinfo {volume} {359}},\ \bibinfo {pages} {4003}
  (\bibinfo {year} {2018})}\BibitemShut {NoStop}%
\bibitem [{\citenamefont {Ma}\ and\ \citenamefont {Shvets}(2016)}]{Ma2016}%
  \BibitemOpen
  \bibfield  {author} {\bibinfo {author} {\bibfnamefont {T.}~\bibnamefont
  {Ma}}\ and\ \bibinfo {author} {\bibfnamefont {G.}~\bibnamefont {Shvets}},\
  }\href {\doibase 10.1088/1367-2630/18/2/025012} {\bibfield  {journal}
  {\bibinfo  {journal} {New. J. Phys}\ }\textbf {\bibinfo {volume} {18}},\
  \bibinfo {pages} {025012} (\bibinfo {year} {2016})}\BibitemShut {NoStop}%
\bibitem [{\citenamefont {Gao}\ \emph {et~al.}(2017)\citenamefont {Gao},
  \citenamefont {Xue}, \citenamefont {Yang}, \citenamefont {Lai}, \citenamefont
  {Yu}, \citenamefont {Lin}, \citenamefont {Chong}, \citenamefont {Shvets},\
  and\ \citenamefont {Zhang}}]{Gao2017}%
  \BibitemOpen
  \bibfield  {author} {\bibinfo {author} {\bibfnamefont {F.}~\bibnamefont
  {Gao}}, \bibinfo {author} {\bibfnamefont {H.}~\bibnamefont {Xue}}, \bibinfo
  {author} {\bibfnamefont {Z.}~\bibnamefont {Yang}}, \bibinfo {author}
  {\bibfnamefont {K.}~\bibnamefont {Lai}}, \bibinfo {author} {\bibfnamefont
  {Y.}~\bibnamefont {Yu}}, \bibinfo {author} {\bibfnamefont {X.}~\bibnamefont
  {Lin}}, \bibinfo {author} {\bibfnamefont {Y.}~\bibnamefont {Chong}}, \bibinfo
  {author} {\bibfnamefont {G.}~\bibnamefont {Shvets}}, \ and\ \bibinfo {author}
  {\bibfnamefont {B.}~\bibnamefont {Zhang}},\ }\href {\doibase
  10.1038/nphys4304
  https://www.nature.com/articles/nphys4304#supplementary-information}
  {\bibfield  {journal} {\bibinfo  {journal} {Nat. Phys.}\ }\textbf {\bibinfo
  {volume} {14}},\ \bibinfo {pages} {140} (\bibinfo {year} {2017})}\BibitemShut
  {NoStop}%
\bibitem [{\citenamefont {He}\ \emph {et~al.}(2019)\citenamefont {He},
  \citenamefont {Liang}, \citenamefont {Yuan}, \citenamefont {Qiu},
  \citenamefont {Chen}, \citenamefont {Zhao},\ and\ \citenamefont
  {Dong}}]{He2019}%
  \BibitemOpen
  \bibfield  {author} {\bibinfo {author} {\bibfnamefont {X.-T.}\ \bibnamefont
  {He}}, \bibinfo {author} {\bibfnamefont {E.-T.}\ \bibnamefont {Liang}},
  \bibinfo {author} {\bibfnamefont {J.-J.}\ \bibnamefont {Yuan}}, \bibinfo
  {author} {\bibfnamefont {H.-Y.}\ \bibnamefont {Qiu}}, \bibinfo {author}
  {\bibfnamefont {X.-D.}\ \bibnamefont {Chen}}, \bibinfo {author}
  {\bibfnamefont {F.-L.}\ \bibnamefont {Zhao}}, \ and\ \bibinfo {author}
  {\bibfnamefont {J.-W.}\ \bibnamefont {Dong}},\ }\href {\doibase
  10.1038/s41467-019-08881-z} {\bibfield  {journal} {\bibinfo  {journal} {Nat.
  Comm.}\ }\textbf {\bibinfo {volume} {10}},\ \bibinfo {pages} {872} (\bibinfo
  {year} {2019})}\BibitemShut {NoStop}%
\bibitem [{\citenamefont {Wu}\ and\ \citenamefont {Hu}(2015)}]{WuHu2015}%
  \BibitemOpen
  \bibfield  {author} {\bibinfo {author} {\bibfnamefont {L.-H.}\ \bibnamefont
  {Wu}}\ and\ \bibinfo {author} {\bibfnamefont {X.}~\bibnamefont {Hu}},\ }\href
  {\doibase 10.1103/PhysRevLett.114.223901} {\bibfield  {journal} {\bibinfo
  {journal} {Phys. Rev. Lett.}\ }\textbf {\bibinfo {volume} {114}},\ \bibinfo
  {pages} {223901} (\bibinfo {year} {2015})}\BibitemShut {NoStop}%
\bibitem [{\citenamefont {Barik}\ \emph {et~al.}(2018)\citenamefont {Barik},
  \citenamefont {Karasahin}, \citenamefont {Flower}, \citenamefont {Cai},
  \citenamefont {Miyake}, \citenamefont {DeGottardi}, \citenamefont {Hafezi},\
  and\ \citenamefont {Waks}}]{Barik2018}%
  \BibitemOpen
  \bibfield  {author} {\bibinfo {author} {\bibfnamefont {S.}~\bibnamefont
  {Barik}}, \bibinfo {author} {\bibfnamefont {A.}~\bibnamefont {Karasahin}},
  \bibinfo {author} {\bibfnamefont {C.}~\bibnamefont {Flower}}, \bibinfo
  {author} {\bibfnamefont {T.}~\bibnamefont {Cai}}, \bibinfo {author}
  {\bibfnamefont {H.}~\bibnamefont {Miyake}}, \bibinfo {author} {\bibfnamefont
  {W.}~\bibnamefont {DeGottardi}}, \bibinfo {author} {\bibfnamefont
  {M.}~\bibnamefont {Hafezi}}, \ and\ \bibinfo {author} {\bibfnamefont
  {E.}~\bibnamefont {Waks}},\ }\href {\doibase 10.1126/science.aaq0327}
  {\bibfield  {journal} {\bibinfo  {journal} {Science}\ }\textbf {\bibinfo
  {volume} {359}},\ \bibinfo {pages} {666} (\bibinfo {year}
  {2018})}\BibitemShut {NoStop}%
\bibitem [{\citenamefont {Mermin}(1979)}]{Mermin1979}%
  \BibitemOpen
  \bibfield  {author} {\bibinfo {author} {\bibfnamefont {N.~D.}\ \bibnamefont
  {Mermin}},\ }\href {\doibase 10.1103/RevModPhys.51.591} {\bibfield  {journal}
  {\bibinfo  {journal} {Rev. Mod. Phys.}\ }\textbf {\bibinfo {volume} {51}},\
  \bibinfo {pages} {591} (\bibinfo {year} {1979})}\BibitemShut {NoStop}%
\bibitem [{\citenamefont {Kosterlitz}(2017)}]{Kosterlitz2017}%
  \BibitemOpen
  \bibfield  {author} {\bibinfo {author} {\bibfnamefont {J.~M.}\ \bibnamefont
  {Kosterlitz}},\ }\href {\doibase 10.1103/RevModPhys.89.040501} {\bibfield
  {journal} {\bibinfo  {journal} {Rev. Mod. Phys.}\ }\textbf {\bibinfo {volume}
  {89}},\ \bibinfo {pages} {040501} (\bibinfo {year} {2017})}\BibitemShut
  {NoStop}%
\bibitem [{\citenamefont {Gonz{\'a}lez}\ \emph {et~al.}(1993)\citenamefont
  {Gonz{\'a}lez}, \citenamefont {Guinea},\ and\ \citenamefont
  {Vozmediano}}]{Gonzalez1993}%
  \BibitemOpen
  \bibfield  {author} {\bibinfo {author} {\bibfnamefont {J.}~\bibnamefont
  {Gonz{\'a}lez}}, \bibinfo {author} {\bibfnamefont {F.}~\bibnamefont
  {Guinea}}, \ and\ \bibinfo {author} {\bibfnamefont {M.~A.~H.}\ \bibnamefont
  {Vozmediano}},\ }\href {\doibase
  https://doi.org/10.1016/0550-3213(93)90009-E} {\bibfield  {journal} {\bibinfo
   {journal} {Nucl. Phys. B}\ }\textbf {\bibinfo {volume} {406}},\ \bibinfo
  {pages} {771} (\bibinfo {year} {1993})}\BibitemShut {NoStop}%
\bibitem [{\citenamefont {Lammert}\ and\ \citenamefont
  {Crespi}(2000)}]{Lammert2000}%
  \BibitemOpen
  \bibfield  {author} {\bibinfo {author} {\bibfnamefont {P.~E.}\ \bibnamefont
  {Lammert}}\ and\ \bibinfo {author} {\bibfnamefont {V.~H.}\ \bibnamefont
  {Crespi}},\ }\href {\doibase 10.1103/PhysRevLett.85.5190} {\bibfield
  {journal} {\bibinfo  {journal} {Phys. Rev. Lett.}\ }\textbf {\bibinfo
  {volume} {85}},\ \bibinfo {pages} {5190} (\bibinfo {year}
  {2000})}\BibitemShut {NoStop}%
\bibitem [{\citenamefont {Vozmediano}\ \emph {et~al.}(2010)\citenamefont
  {Vozmediano}, \citenamefont {Katsnelson},\ and\ \citenamefont
  {Guinea}}]{Vozmediano2010}%
  \BibitemOpen
  \bibfield  {author} {\bibinfo {author} {\bibfnamefont {M.~A.~H.}\
  \bibnamefont {Vozmediano}}, \bibinfo {author} {\bibfnamefont {M.~I.}\
  \bibnamefont {Katsnelson}}, \ and\ \bibinfo {author} {\bibfnamefont
  {F.}~\bibnamefont {Guinea}},\ }\href {\doibase
  https://doi.org/10.1016/j.physrep.2010.07.003} {\bibfield  {journal}
  {\bibinfo  {journal} {Phys. Rep.}\ }\textbf {\bibinfo {volume} {496}},\
  \bibinfo {pages} {109} (\bibinfo {year} {2010})}\BibitemShut {NoStop}%
\bibitem [{\citenamefont {Kotakoski}\ \emph {et~al.}(2011)\citenamefont
  {Kotakoski}, \citenamefont {Krasheninnikov}, \citenamefont {Kaiser},\ and\
  \citenamefont {Meyer}}]{Kotakoski2011}%
  \BibitemOpen
  \bibfield  {author} {\bibinfo {author} {\bibfnamefont {J.}~\bibnamefont
  {Kotakoski}}, \bibinfo {author} {\bibfnamefont {A.~V.}\ \bibnamefont
  {Krasheninnikov}}, \bibinfo {author} {\bibfnamefont {U.}~\bibnamefont
  {Kaiser}}, \ and\ \bibinfo {author} {\bibfnamefont {J.~C.}\ \bibnamefont
  {Meyer}},\ }\href {\doibase 10.1103/PhysRevLett.106.105505} {\bibfield
  {journal} {\bibinfo  {journal} {Phys. Rev. Lett.}\ }\textbf {\bibinfo
  {volume} {106}},\ \bibinfo {pages} {105505} (\bibinfo {year}
  {2011})}\BibitemShut {NoStop}%
\bibitem [{\citenamefont {de~Souza}\ \emph {et~al.}(2014)\citenamefont
  {de~Souza}, \citenamefont {de~Lima~Ribeiro},\ and\ \citenamefont
  {Furtado}}]{deSouza2014}%
  \BibitemOpen
  \bibfield  {author} {\bibinfo {author} {\bibfnamefont {J.}~\bibnamefont
  {de~Souza}}, \bibinfo {author} {\bibfnamefont {C.}~\bibnamefont
  {de~Lima~Ribeiro}}, \ and\ \bibinfo {author} {\bibfnamefont {C.}~\bibnamefont
  {Furtado}},\ }\href {\doibase 10.1016/j.physleta.2014.05.053} {\bibfield
  {journal} {\bibinfo  {journal} {Physics Letters A}\ }\textbf {\bibinfo
  {volume} {378}},\ \bibinfo {pages} {2317–2324} (\bibinfo {year}
  {2014})}\BibitemShut {NoStop}%
\bibitem [{\citenamefont {Yazyev}\ and\ \citenamefont
  {Louie}(2010)}]{Yazyev2010}%
  \BibitemOpen
  \bibfield  {author} {\bibinfo {author} {\bibfnamefont {O.~V.}\ \bibnamefont
  {Yazyev}}\ and\ \bibinfo {author} {\bibfnamefont {S.~G.}\ \bibnamefont
  {Louie}},\ }\href {\doibase 10.1038/nmat2830} {\bibfield  {journal} {\bibinfo
   {journal} {Nat. Mater.}\ }\textbf {\bibinfo {volume} {9}},\ \bibinfo {pages}
  {806} (\bibinfo {year} {2010})}\BibitemShut {NoStop}%
\bibitem [{\citenamefont {Lahiri}\ \emph {et~al.}(2010)\citenamefont {Lahiri},
  \citenamefont {Lin}, \citenamefont {Bozkurt}, \citenamefont {Oleynik},\ and\
  \citenamefont {Batzill}}]{Lahiri2010}%
  \BibitemOpen
  \bibfield  {author} {\bibinfo {author} {\bibfnamefont {J.}~\bibnamefont
  {Lahiri}}, \bibinfo {author} {\bibfnamefont {Y.}~\bibnamefont {Lin}},
  \bibinfo {author} {\bibfnamefont {P.}~\bibnamefont {Bozkurt}}, \bibinfo
  {author} {\bibfnamefont {I.~I.}\ \bibnamefont {Oleynik}}, \ and\ \bibinfo
  {author} {\bibfnamefont {M.}~\bibnamefont {Batzill}},\ }\href {\doibase
  10.1038/nnano.2010.53} {\bibfield  {journal} {\bibinfo  {journal} {Nat.
  Nanotech.}\ }\textbf {\bibinfo {volume} {5}},\ \bibinfo {pages} {326}
  (\bibinfo {year} {2010})}\BibitemShut {NoStop}%
\bibitem [{\citenamefont {Huang}\ \emph {et~al.}(2011)\citenamefont {Huang},
  \citenamefont {Ruiz-Vargas}, \citenamefont {van~der Zande}, \citenamefont
  {Whitney}, \citenamefont {Levendorf}, \citenamefont {Kevek}, \citenamefont
  {Garg}, \citenamefont {Alden}, \citenamefont {Hustedt}, \citenamefont {Zhu},
  \citenamefont {Park}, \citenamefont {McEuen},\ and\ \citenamefont
  {Muller}}]{Huang2011}%
  \BibitemOpen
  \bibfield  {author} {\bibinfo {author} {\bibfnamefont {P.~Y.}\ \bibnamefont
  {Huang}}, \bibinfo {author} {\bibfnamefont {C.~S.}\ \bibnamefont
  {Ruiz-Vargas}}, \bibinfo {author} {\bibfnamefont {A.~M.}\ \bibnamefont
  {van~der Zande}}, \bibinfo {author} {\bibfnamefont {W.~S.}\ \bibnamefont
  {Whitney}}, \bibinfo {author} {\bibfnamefont {M.~P.}\ \bibnamefont
  {Levendorf}}, \bibinfo {author} {\bibfnamefont {J.~W.}\ \bibnamefont
  {Kevek}}, \bibinfo {author} {\bibfnamefont {S.}~\bibnamefont {Garg}},
  \bibinfo {author} {\bibfnamefont {J.~S.}\ \bibnamefont {Alden}}, \bibinfo
  {author} {\bibfnamefont {C.~J.}\ \bibnamefont {Hustedt}}, \bibinfo {author}
  {\bibfnamefont {Y.}~\bibnamefont {Zhu}}, \bibinfo {author} {\bibfnamefont
  {J.}~\bibnamefont {Park}}, \bibinfo {author} {\bibfnamefont {P.~L.}\
  \bibnamefont {McEuen}}, \ and\ \bibinfo {author} {\bibfnamefont {D.~A.}\
  \bibnamefont {Muller}},\ }\href {\doibase 10.1038/nature09718} {\bibfield
  {journal} {\bibinfo  {journal} {Nature}\ }\textbf {\bibinfo {volume} {469}},\
  \bibinfo {pages} {389} (\bibinfo {year} {2011})}\BibitemShut {NoStop}%
\bibitem [{\citenamefont {Warner}\ \emph {et~al.}(2012)\citenamefont {Warner},
  \citenamefont {Margine}, \citenamefont {Mukai}, \citenamefont {Robertson},
  \citenamefont {Giustino},\ and\ \citenamefont {Kirkland}}]{Warner2012}%
  \BibitemOpen
  \bibfield  {author} {\bibinfo {author} {\bibfnamefont {J.~H.}\ \bibnamefont
  {Warner}}, \bibinfo {author} {\bibfnamefont {E.~R.}\ \bibnamefont {Margine}},
  \bibinfo {author} {\bibfnamefont {M.}~\bibnamefont {Mukai}}, \bibinfo
  {author} {\bibfnamefont {A.~W.}\ \bibnamefont {Robertson}}, \bibinfo {author}
  {\bibfnamefont {F.}~\bibnamefont {Giustino}}, \ and\ \bibinfo {author}
  {\bibfnamefont {A.~I.}\ \bibnamefont {Kirkland}},\ }\href {\doibase
  10.1126/science.1217529} {\bibfield  {journal} {\bibinfo  {journal}
  {Science}\ }\textbf {\bibinfo {volume} {337}},\ \bibinfo {pages} {209}
  (\bibinfo {year} {2012})}\BibitemShut {NoStop}%
\bibitem [{\citenamefont {Plotnik}\ \emph {et~al.}(2014)\citenamefont
  {Plotnik}, \citenamefont {Rechtsman}, \citenamefont {Song}, \citenamefont
  {Heinrich}, \citenamefont {Zeuner}, \citenamefont {Nolte}, \citenamefont
  {Lumer}, \citenamefont {Malkova}, \citenamefont {Xu}, \citenamefont
  {Szameit},\ and\ \citenamefont {et~al.}}]{Plotnik2014}%
  \BibitemOpen
  \bibfield  {author} {\bibinfo {author} {\bibfnamefont {Y.}~\bibnamefont
  {Plotnik}}, \bibinfo {author} {\bibfnamefont {M.~C.}\ \bibnamefont
  {Rechtsman}}, \bibinfo {author} {\bibfnamefont {D.}~\bibnamefont {Song}},
  \bibinfo {author} {\bibfnamefont {M.}~\bibnamefont {Heinrich}}, \bibinfo
  {author} {\bibfnamefont {J.~M.}\ \bibnamefont {Zeuner}}, \bibinfo {author}
  {\bibfnamefont {S.}~\bibnamefont {Nolte}}, \bibinfo {author} {\bibfnamefont
  {Y.}~\bibnamefont {Lumer}}, \bibinfo {author} {\bibfnamefont
  {N.}~\bibnamefont {Malkova}}, \bibinfo {author} {\bibfnamefont
  {J.}~\bibnamefont {Xu}}, \bibinfo {author} {\bibfnamefont {A.}~\bibnamefont
  {Szameit}}, \ and\ \bibinfo {author} {\bibnamefont {et~al.}},\ }\href
  {\doibase 10.1038/nmat3783} {\bibfield  {journal} {\bibinfo  {journal}
  {Nature Materials}\ }\textbf {\bibinfo {volume} {13}},\ \bibinfo {pages}
  {57–62} (\bibinfo {year} {2014})}\BibitemShut {NoStop}%
\bibitem [{\citenamefont {R\"uegg}\ and\ \citenamefont
  {Lin}(2013)}]{Ruegg2013}%
  \BibitemOpen
  \bibfield  {author} {\bibinfo {author} {\bibfnamefont {A.}~\bibnamefont
  {R\"uegg}}\ and\ \bibinfo {author} {\bibfnamefont {C.}~\bibnamefont {Lin}},\
  }\href {\doibase 10.1103/PhysRevLett.110.046401} {\bibfield  {journal}
  {\bibinfo  {journal} {Phys. Rev. Lett.}\ }\textbf {\bibinfo {volume} {110}},\
  \bibinfo {pages} {046401} (\bibinfo {year} {2013})}\BibitemShut {NoStop}%
\bibitem [{\citenamefont {Miyazaki}\ \emph {et~al.}(2003)\citenamefont
  {Miyazaki}, \citenamefont {Hase}, \citenamefont {Miyazaki}, \citenamefont
  {Kurokawa},\ and\ \citenamefont {Shinya}}]{Miyazaki2003}%
  \BibitemOpen
  \bibfield  {author} {\bibinfo {author} {\bibfnamefont {H.}~\bibnamefont
  {Miyazaki}}, \bibinfo {author} {\bibfnamefont {M.}~\bibnamefont {Hase}},
  \bibinfo {author} {\bibfnamefont {H.~T.}\ \bibnamefont {Miyazaki}}, \bibinfo
  {author} {\bibfnamefont {Y.}~\bibnamefont {Kurokawa}}, \ and\ \bibinfo
  {author} {\bibfnamefont {N.}~\bibnamefont {Shinya}},\ }\href {\doibase
  10.1103/PhysRevB.67.235109} {\bibfield  {journal} {\bibinfo  {journal} {Phys.
  Rev. B}\ }\textbf {\bibinfo {volume} {67}},\ \bibinfo {pages} {235109}
  (\bibinfo {year} {2003})}\BibitemShut {NoStop}%
\bibitem [{\citenamefont {Edagawa}\ \emph {et~al.}(2008)\citenamefont
  {Edagawa}, \citenamefont {Kanoko},\ and\ \citenamefont
  {Notomi}}]{Edagawa2008}%
  \BibitemOpen
  \bibfield  {author} {\bibinfo {author} {\bibfnamefont {K.}~\bibnamefont
  {Edagawa}}, \bibinfo {author} {\bibfnamefont {S.}~\bibnamefont {Kanoko}}, \
  and\ \bibinfo {author} {\bibfnamefont {M.}~\bibnamefont {Notomi}},\ }\href
  {\doibase 10.1103/PhysRevLett.100.013901} {\bibfield  {journal} {\bibinfo
  {journal} {Phys. Rev. Lett.}\ }\textbf {\bibinfo {volume} {100}},\ \bibinfo
  {pages} {013901} (\bibinfo {year} {2008})}\BibitemShut {NoStop}%
\bibitem [{\citenamefont {Florescu}\ \emph {et~al.}(2009)\citenamefont
  {Florescu}, \citenamefont {Torquato},\ and\ \citenamefont
  {Steinhardt}}]{Florescu2009}%
  \BibitemOpen
  \bibfield  {author} {\bibinfo {author} {\bibfnamefont {M.}~\bibnamefont
  {Florescu}}, \bibinfo {author} {\bibfnamefont {S.}~\bibnamefont {Torquato}},
  \ and\ \bibinfo {author} {\bibfnamefont {P.~J.}\ \bibnamefont {Steinhardt}},\
  }\href {\doibase 10.1073/pnas.0907744106} {\bibfield  {journal} {\bibinfo
  {journal} {Proc. Natl. Acad. Sci.}\ }\textbf {\bibinfo {volume} {106}},\
  \bibinfo {pages} {20658–20663} (\bibinfo {year} {2009})}\BibitemShut
  {NoStop}%
\bibitem [{\citenamefont {Yang}\ \emph {et~al.}(2010)\citenamefont {Yang},
  \citenamefont {Schreck}, \citenamefont {Noh}, \citenamefont {Liew},
  \citenamefont {Guy}, \citenamefont {O'Hern},\ and\ \citenamefont
  {Cao}}]{Yang2010}%
  \BibitemOpen
  \bibfield  {author} {\bibinfo {author} {\bibfnamefont {J.-K.}\ \bibnamefont
  {Yang}}, \bibinfo {author} {\bibfnamefont {C.}~\bibnamefont {Schreck}},
  \bibinfo {author} {\bibfnamefont {H.}~\bibnamefont {Noh}}, \bibinfo {author}
  {\bibfnamefont {S.-F.}\ \bibnamefont {Liew}}, \bibinfo {author}
  {\bibfnamefont {M.~I.}\ \bibnamefont {Guy}}, \bibinfo {author} {\bibfnamefont
  {C.~S.}\ \bibnamefont {O'Hern}}, \ and\ \bibinfo {author} {\bibfnamefont
  {H.}~\bibnamefont {Cao}},\ }\href {\doibase 10.1103/PhysRevA.82.053838}
  {\bibfield  {journal} {\bibinfo  {journal} {Phys. Rev. A}\ }\textbf {\bibinfo
  {volume} {82}},\ \bibinfo {pages} {053838} (\bibinfo {year}
  {2010})}\BibitemShut {NoStop}%
\bibitem [{\citenamefont {Imagawa}\ \emph {et~al.}(2010)\citenamefont
  {Imagawa}, \citenamefont {Edagawa}, \citenamefont {Morita}, \citenamefont
  {Niino}, \citenamefont {Kagawa},\ and\ \citenamefont {Notomi}}]{Imagawa2010}%
  \BibitemOpen
  \bibfield  {author} {\bibinfo {author} {\bibfnamefont {S.}~\bibnamefont
  {Imagawa}}, \bibinfo {author} {\bibfnamefont {K.}~\bibnamefont {Edagawa}},
  \bibinfo {author} {\bibfnamefont {K.}~\bibnamefont {Morita}}, \bibinfo
  {author} {\bibfnamefont {T.}~\bibnamefont {Niino}}, \bibinfo {author}
  {\bibfnamefont {Y.}~\bibnamefont {Kagawa}}, \ and\ \bibinfo {author}
  {\bibfnamefont {M.}~\bibnamefont {Notomi}},\ }\href {\doibase
  10.1103/PhysRevB.82.115116} {\bibfield  {journal} {\bibinfo  {journal} {Phys.
  Rev. B}\ }\textbf {\bibinfo {volume} {82}},\ \bibinfo {pages} {115116}
  (\bibinfo {year} {2010})}\BibitemShut {NoStop}%
\bibitem [{\citenamefont {Man}\ \emph {et~al.}(2013)\citenamefont {Man},
  \citenamefont {Florescu}, \citenamefont {Williamson}, \citenamefont {He},
  \citenamefont {Hashemizad}, \citenamefont {Leung}, \citenamefont {Liner},
  \citenamefont {Torquato}, \citenamefont {Chaikin},\ and\ \citenamefont
  {Steinhardt}}]{Man2013}%
  \BibitemOpen
  \bibfield  {author} {\bibinfo {author} {\bibfnamefont {W.}~\bibnamefont
  {Man}}, \bibinfo {author} {\bibfnamefont {M.}~\bibnamefont {Florescu}},
  \bibinfo {author} {\bibfnamefont {E.~P.}\ \bibnamefont {Williamson}},
  \bibinfo {author} {\bibfnamefont {Y.}~\bibnamefont {He}}, \bibinfo {author}
  {\bibfnamefont {S.~R.}\ \bibnamefont {Hashemizad}}, \bibinfo {author}
  {\bibfnamefont {B.~Y.~C.}\ \bibnamefont {Leung}}, \bibinfo {author}
  {\bibfnamefont {D.~R.}\ \bibnamefont {Liner}}, \bibinfo {author}
  {\bibfnamefont {S.}~\bibnamefont {Torquato}}, \bibinfo {author}
  {\bibfnamefont {P.~M.}\ \bibnamefont {Chaikin}}, \ and\ \bibinfo {author}
  {\bibfnamefont {P.~J.}\ \bibnamefont {Steinhardt}},\ }\href {\doibase
  10.1073/pnas.1307879110} {\bibfield  {journal} {\bibinfo  {journal} {Proc.
  Natl. Acad. Sci.}\ }\textbf {\bibinfo {volume} {110}},\ \bibinfo {pages}
  {15886–15891} (\bibinfo {year} {2013})}\BibitemShut {NoStop}%
\bibitem [{\citenamefont {Florescu}\ \emph {et~al.}(2013)\citenamefont
  {Florescu}, \citenamefont {Steinhardt},\ and\ \citenamefont
  {Torquato}}]{Florescu2013}%
  \BibitemOpen
  \bibfield  {author} {\bibinfo {author} {\bibfnamefont {M.}~\bibnamefont
  {Florescu}}, \bibinfo {author} {\bibfnamefont {P.~J.}\ \bibnamefont
  {Steinhardt}}, \ and\ \bibinfo {author} {\bibfnamefont {S.}~\bibnamefont
  {Torquato}},\ }\href {\doibase 10.1103/PhysRevB.87.165116} {\bibfield
  {journal} {\bibinfo  {journal} {Phys. Rev. B}\ }\textbf {\bibinfo {volume}
  {87}},\ \bibinfo {pages} {165116} (\bibinfo {year} {2013})}\BibitemShut
  {NoStop}%
\bibitem [{\citenamefont {Kramer}\ and\ \citenamefont
  {MacKinnon}(1993)}]{Kramer1993}%
  \BibitemOpen
  \bibfield  {author} {\bibinfo {author} {\bibfnamefont {B.}~\bibnamefont
  {Kramer}}\ and\ \bibinfo {author} {\bibfnamefont {A.}~\bibnamefont
  {MacKinnon}},\ }\href {\doibase 10.1088/0034-4885/56/12/001} {\bibfield
  {journal} {\bibinfo  {journal} {Rep. Prog. Phys.}\ }\textbf {\bibinfo
  {volume} {56}},\ \bibinfo {pages} {1469–1564} (\bibinfo {year}
  {1993})}\BibitemShut {NoStop}%
\bibitem [{\citenamefont {Bandres}\ \emph {et~al.}(2016)\citenamefont
  {Bandres}, \citenamefont {Rechtsman},\ and\ \citenamefont
  {Segev}}]{Bandres2016}%
  \BibitemOpen
  \bibfield  {author} {\bibinfo {author} {\bibfnamefont {M.~A.}\ \bibnamefont
  {Bandres}}, \bibinfo {author} {\bibfnamefont {M.~C.}\ \bibnamefont
  {Rechtsman}}, \ and\ \bibinfo {author} {\bibfnamefont {M.}~\bibnamefont
  {Segev}},\ }\href {\doibase 10.1103/PhysRevX.6.011016} {\bibfield  {journal}
  {\bibinfo  {journal} {Phys. Rev. X}\ }\textbf {\bibinfo {volume} {6}},\
  \bibinfo {pages} {011016} (\bibinfo {year} {2016})}\BibitemShut {NoStop}%
\bibitem [{\citenamefont {Mitchell}\ \emph {et~al.}(2018)\citenamefont
  {Mitchell}, \citenamefont {Nash}, \citenamefont {Hexner}, \citenamefont
  {Turner},\ and\ \citenamefont {Irvine}}]{Mitchell2018}%
  \BibitemOpen
  \bibfield  {author} {\bibinfo {author} {\bibfnamefont {N.~P.}\ \bibnamefont
  {Mitchell}}, \bibinfo {author} {\bibfnamefont {L.~M.}\ \bibnamefont {Nash}},
  \bibinfo {author} {\bibfnamefont {D.}~\bibnamefont {Hexner}}, \bibinfo
  {author} {\bibfnamefont {A.~M.}\ \bibnamefont {Turner}}, \ and\ \bibinfo
  {author} {\bibfnamefont {W.~T.~M.}\ \bibnamefont {Irvine}},\ }\href {\doibase
  10.1038/s41567-017-0024-5} {\bibfield  {journal} {\bibinfo  {journal} {Nat.
  Phys.}\ }\textbf {\bibinfo {volume} {14}},\ \bibinfo {pages} {380–385}
  (\bibinfo {year} {2018})}\BibitemShut {NoStop}%
\bibitem [{\citenamefont {Lammert}\ and\ \citenamefont
  {Crespi}(2004)}]{Lammert2004}%
  \BibitemOpen
  \bibfield  {author} {\bibinfo {author} {\bibfnamefont {P.~E.}\ \bibnamefont
  {Lammert}}\ and\ \bibinfo {author} {\bibfnamefont {V.~H.}\ \bibnamefont
  {Crespi}},\ }\href {\doibase 10.1103/PhysRevB.69.035406} {\bibfield
  {journal} {\bibinfo  {journal} {Phys. Rev. B}\ }\textbf {\bibinfo {volume}
  {69}},\ \bibinfo {pages} {035406} (\bibinfo {year} {2004})}\BibitemShut
  {NoStop}%
\bibitem [{\citenamefont {Plimpton}(1995)}]{lammps}%
  \BibitemOpen
  \bibfield  {author} {\bibinfo {author} {\bibfnamefont {S.}~\bibnamefont
  {Plimpton}},\ }\href@noop {} {\bibfield  {journal} {\bibinfo  {journal} {J.
  Comp. Phys.}\ }\textbf {\bibinfo {volume} {117}},\ \bibinfo {pages} {1}
  (\bibinfo {year} {1995})}\BibitemShut {NoStop}%
\bibitem [{\citenamefont {Mansha}\ \emph {et~al.}(2016)\citenamefont {Mansha},
  \citenamefont {Zeng}, \citenamefont {Wang},\ and\ \citenamefont
  {Chong}}]{Mansha2016}%
  \BibitemOpen
  \bibfield  {author} {\bibinfo {author} {\bibfnamefont {S.}~\bibnamefont
  {Mansha}}, \bibinfo {author} {\bibfnamefont {Y.}~\bibnamefont {Zeng}},
  \bibinfo {author} {\bibfnamefont {Q.~J.}\ \bibnamefont {Wang}}, \ and\
  \bibinfo {author} {\bibfnamefont {Y.~D.}\ \bibnamefont {Chong}},\ }\href
  {\doibase 10.1364/OE.24.004890} {\bibfield  {journal} {\bibinfo  {journal}
  {Opt. Express}\ }\textbf {\bibinfo {volume} {24}},\ \bibinfo {pages} {4890}
  (\bibinfo {year} {2016})}\BibitemShut {NoStop}%
\bibitem [{\citenamefont {Cheng}\ \emph {et~al.}(2016)\citenamefont {Cheng},
  \citenamefont {Jouvaud}, \citenamefont {Ni}, \citenamefont {Mousavi},
  \citenamefont {Genack},\ and\ \citenamefont {Khanikaev}}]{Cheng2016}%
  \BibitemOpen
  \bibfield  {author} {\bibinfo {author} {\bibfnamefont {X.}~\bibnamefont
  {Cheng}}, \bibinfo {author} {\bibfnamefont {C.}~\bibnamefont {Jouvaud}},
  \bibinfo {author} {\bibfnamefont {X.}~\bibnamefont {Ni}}, \bibinfo {author}
  {\bibfnamefont {S.~H.}\ \bibnamefont {Mousavi}}, \bibinfo {author}
  {\bibfnamefont {A.~Z.}\ \bibnamefont {Genack}}, \ and\ \bibinfo {author}
  {\bibfnamefont {A.~B.}\ \bibnamefont {Khanikaev}},\ }\href {\doibase
  10.1038/nmat4573} {\bibfield  {journal} {\bibinfo  {journal} {Nat. Mater.}\
  }\textbf {\bibinfo {volume} {15}},\ \bibinfo {pages} {542–548} (\bibinfo
  {year} {2016})}\BibitemShut {NoStop}%
\bibitem [{\citenamefont {Noh}\ \emph {et~al.}(2018)\citenamefont {Noh},
  \citenamefont {Benalcazar}, \citenamefont {Huang}, \citenamefont {Collins},
  \citenamefont {Chen}, \citenamefont {Hughes},\ and\ \citenamefont
  {Rechtsman}}]{Noh2018}%
  \BibitemOpen
  \bibfield  {author} {\bibinfo {author} {\bibfnamefont {J.}~\bibnamefont
  {Noh}}, \bibinfo {author} {\bibfnamefont {W.~A.}\ \bibnamefont {Benalcazar}},
  \bibinfo {author} {\bibfnamefont {S.}~\bibnamefont {Huang}}, \bibinfo
  {author} {\bibfnamefont {M.~J.}\ \bibnamefont {Collins}}, \bibinfo {author}
  {\bibfnamefont {K.~P.}\ \bibnamefont {Chen}}, \bibinfo {author}
  {\bibfnamefont {T.~L.}\ \bibnamefont {Hughes}}, \ and\ \bibinfo {author}
  {\bibfnamefont {M.~C.}\ \bibnamefont {Rechtsman}},\ }\href {\doibase
  10.1038/s41566-018-0179-3} {\bibfield  {journal} {\bibinfo  {journal} {Nat.
  Photon}\ }\textbf {\bibinfo {volume} {12}},\ \bibinfo {pages} {408} (\bibinfo
  {year} {2018})}\BibitemShut {NoStop}%
\bibitem [{\citenamefont {Sitenko}\ and\ \citenamefont
  {Vlasii}(2007)}]{Sitenko2007}%
  \BibitemOpen
  \bibfield  {author} {\bibinfo {author} {\bibfnamefont {Y.~A.}\ \bibnamefont
  {Sitenko}}\ and\ \bibinfo {author} {\bibfnamefont {N.~D.}\ \bibnamefont
  {Vlasii}},\ }\href {\doibase https://doi.org/10.1016/j.nuclphysb.2007.06.001}
  {\bibfield  {journal} {\bibinfo  {journal} {Nucl. Phys. B}\ }\textbf
  {\bibinfo {volume} {787}},\ \bibinfo {pages} {241} (\bibinfo {year}
  {2007})}\BibitemShut {NoStop}%
\bibitem [{\citenamefont {Cortijo}\ and\ \citenamefont
  {Vozmediano}(2007)}]{Cortijo2007}%
  \BibitemOpen
  \bibfield  {author} {\bibinfo {author} {\bibfnamefont {A.}~\bibnamefont
  {Cortijo}}\ and\ \bibinfo {author} {\bibfnamefont {M.~A.}\ \bibnamefont
  {Vozmediano}},\ }\href {https://doi.org/10.1140/epjst/e2007-00228-2}
  {\bibfield  {journal} {\bibinfo  {journal} {Eur. Phys. J. Spec. Top.}\
  }\textbf {\bibinfo {volume} {148}},\ \bibinfo {pages} {83} (\bibinfo {year}
  {2007})}\BibitemShut {NoStop}%
\bibitem [{\citenamefont {Jeong}\ \emph {et~al.}(2008)\citenamefont {Jeong},
  \citenamefont {Ihm},\ and\ \citenamefont {Lee}}]{Jeong2008}%
  \BibitemOpen
  \bibfield  {author} {\bibinfo {author} {\bibfnamefont {B.~W.}\ \bibnamefont
  {Jeong}}, \bibinfo {author} {\bibfnamefont {J.}~\bibnamefont {Ihm}}, \ and\
  \bibinfo {author} {\bibfnamefont {G.-D.}\ \bibnamefont {Lee}},\ }\href
  {\doibase 10.1103/PhysRevB.78.165403} {\bibfield  {journal} {\bibinfo
  {journal} {Phys. Rev. B}\ }\textbf {\bibinfo {volume} {78}},\ \bibinfo
  {pages} {165403} (\bibinfo {year} {2008})}\BibitemShut {NoStop}%
\bibitem [{\citenamefont {Wei}\ \emph {et~al.}(2012)\citenamefont {Wei},
  \citenamefont {Wu}, \citenamefont {Yin}, \citenamefont {Shi}, \citenamefont
  {Yang},\ and\ \citenamefont {Dresselhaus}}]{wei2012}%
  \BibitemOpen
  \bibfield  {author} {\bibinfo {author} {\bibfnamefont {Y.}~\bibnamefont
  {Wei}}, \bibinfo {author} {\bibfnamefont {J.}~\bibnamefont {Wu}}, \bibinfo
  {author} {\bibfnamefont {H.}~\bibnamefont {Yin}}, \bibinfo {author}
  {\bibfnamefont {X.}~\bibnamefont {Shi}}, \bibinfo {author} {\bibfnamefont
  {R.}~\bibnamefont {Yang}}, \ and\ \bibinfo {author} {\bibfnamefont
  {M.}~\bibnamefont {Dresselhaus}},\ }\href {\doibase 10.1038/nmat3370
  https://www.nature.com/articles/nmat3370#supplementary-information}
  {\bibfield  {journal} {\bibinfo  {journal} {Nat. Mater}\ }\textbf {\bibinfo
  {volume} {11}},\ \bibinfo {pages} {759} (\bibinfo {year} {2012})}\BibitemShut
  {NoStop}%
\end{thebibliography}%

%% \clearpage

%% \begin{widetext}

%% \makeatletter 
%% \renewcommand{\theequation}{S\arabic{equation}}
%% \makeatother
%% \setcounter{equation}{0}

%% \makeatletter 
%% \renewcommand{\thefigure}{S\@arabic\c@figure}
%% \makeatother
%% \setcounter{figure}{0}

%% \begin{center}
%% {\Large Supplemental  Material For \\"Observation of Protected Photonic Edge States Induced By\\Real-Space Topological Lattice Defectsl"}
%% \end{center}

\end{document}